\documentclass[twocolumn,onecolappendix]{aastex6}

\usepackage{lipsum}
\usepackage{amsmath}
\usepackage{longtable}

\begin{document}


\title{Thermo-compositional diabatic convection in the atmospheres of brown dwarfs and in Earth's atmosphere and oceans}

%% Use \author, \affil, plus the \and command to format author and affiliation 
%% information.  If done correctly the peer review system will be able to
%% automatically put the author and affiliation information from the manuscript
%% and save the corresponding author the trouble of entering it by hand.
%%
%% The \affil should be used to document primary affiliations and the
%% \altaffil should be used for secondary affiliations, titles, or email.

%% Authors with the same affiliation can be grouped in a single

\author{ 
P. Tremblin     \altaffilmark{1},
T. Padioleau    \altaffilmark{1},
M. Phillips     \altaffilmark{2},
G. Chabrier     \altaffilmark{2,3},
I. Baraffe      \altaffilmark{2,3},
S. Fromang      \altaffilmark{4},
E. Audit        \altaffilmark{1},
H. Bloch        \altaffilmark{1},
A. J. Burgasser \altaffilmark{5},
B. Drummond     \altaffilmark{2},
M. Gonz\'alez   \altaffilmark{4},
P. Kestener     \altaffilmark{1},
S. Kokh         \altaffilmark{6},
P.-O. Lagage    \altaffilmark{4}, and
M. Stauffert    \altaffilmark{1}
       }

\altaffiltext{1}{
Maison de la Simulation, CEA, CNRS, Univ. Paris-Sud, UVSQ, Université Paris-Saclay, 91191 
Gif-sur-Yvette, France} 

\altaffiltext{2}{
  Astrophysics Group, University of Exeter, EX4 4QL Exeter, UK}
  
\altaffiltext{3}{
  Ecole Normale Sup\'erieure de Lyon, CRAL, UMR CNRS 5574, 69364 Lyon  Cedex 07, France}

\altaffiltext{4}{
  AIM, CEA, CNRS, Universit\'e Paris-Saclay, Universit\'e Paris Diderot, Sorbonne Paris Cité, UMR7158 F-91191 Gif-sur-Yvette, France}
  
\altaffiltext{5}{
UC San Diego, M/C 0424, 9500 Gilman Drive, La Jolla, CA 92093, USA}

\altaffiltext{6}{
DEN/DANS/DM2S/STMF, CEA Saclay, 91191 Gif-sur-Yvette, France}

\email{pascal.tremblin@cea.fr}

\begin{abstract}
  By generalizing the theory of convection to any type of thermal and compositional source terms (diabatic processes), we show that thermohaline convection in Earth oceans, fingering convection in stellar atmospheres, and moist convection in Earth atmosphere are deriving  from the same general diabatic convective instability. We show also that "radiative convection" triggered by CO/CH4 transition with radiative transfer in the atmospheres of brown dwarfs is analog to moist and thermohaline convection.
We derive a generalization of the mixing length theory to include the effect of source terms in 1D codes. We show that CO/CH$_4$ ``radiative'' convection could significantly reduce the temperature gradient in the atmospheres of brown dwarfs similarly to moist convection in Earth atmosphere thus possibly explaining the reddening in brown-dwarf spectra. By using idealized two-dimensional hydrodynamic simulations in the Ledoux unstable regime, we show that compositional source terms can indeed provoke a reduction of the temperature gradien. The L/T transition could be explained by a bifurcation between the adiabatic and diabatic convective transports and could be seen as a giant cooling crisis: an analog of the boiling crisis in liquid/steam-water convective flows.

This mechanism with other chemical transitions could be present in many giant and earth-like exoplanets. The study of the impact of different parameters (effective temperature, compositional changes) on CO/CH$_4$ radiative convection and the analogy with Earth moist and thermohaline convection is opening the possibility to use brown dwarfs to better understand some aspects of the physics at play in the climate of our own planet.
\end{abstract}

  
\section{Introduction}
Rayleigh-Benard convection is a well-known physical process occurring in many physical and astrophysical systems such as the interior and atmosphere of stars, brown dwarfs, and gaseous planets. This process has been widely studied with theory, numerical simulations, and experiments leading to a relatively clear understanding of this mechanism \citep[see e.g.][for a review]{Bodenschatz_2000}. When viscosity is neglected, the instability criterion reduces to the standard Schwarzschild or Ledoux criterion for convection.

On the contrary the inclusion in the theory of compositional and thermal source terms (diabatic processes) has remained very limited up to now. The thermal and compositional diffusion processes have been considered leading to the extension of the theory to double-diffusive convection known as thermohaline convection in Earth oceans with salt diffusion \citep{Stern_1960}, and known as fingering convection in the interior of stars with the diffusion of heavy elements \citep[in case of a stabilizing temperature gradient and a destabilizing molecular weight gradient,][]{Ulrich_1972}. Theoretical developments for the study of the linear phase of this instability can be found in e.g. \citet{1966PASJ...18..374K,1969JFM....37..289B}. Pioneering works on the modeling side have been realized in the past decades \citep{Denissenkov_2010,Stellmach_2011,Traxler_2011,Brown_2013}, but the interest for these subtle and complicated processes has remained limited in the astrophysical community because of the difficulty to constrain its effect in observations and its potential inefficiency in some situations \citep[e.g.][]{Wachlin_2014}.

In a series of papers, we have recently proposed that an analog of fingering convection could occur in the atmosphere of brown dwarfs and extrasolar giant planets \citet{tremblin:2015aa,tremblin:2016aa,2017ApJ...850...46T}. We have demonstrated that numerous observations could be reproduced with a relatively simple model assuming that this process leads to a reduction of the temperature gradient (as a function of pressure) in the atmosphere. The thermal and compositional diffusion processes would be replaced by radiative transfer and chemical reactions such as the conversion between CO and CH$_4$ in the atmospheres of brown dwarfs. It is clear though that convection theory needs to be extended to consider these types of source terms that are not strictly speaking diffusive processes. 

In this paper we propose such an extension of the theory of convection that is able to take into account any type of source terms. This extension is opening a new paradigm that can encompass numerous diabatic convective systems such as thermohaline and fingering convection but also moist convection occurring in the Earth atmosphere and CO/CH$_4$ ``radiative'' convection in the atmosphere of brown dwarfs. In Sect.~\ref{sect:linear}, we first develop the theoretical framework in the linear regime and show that all these convective processes are deriving from the same instability criterion. In Sect.~\ref{sect:non-linear}, we study the non-linear regime of the instability by using mean field approach to develop a mixing length theory that includes the convective thermal and compositional transport.
Finally, in Sect.~\ref{sect:models}, we use this new mixing length theory in a 1D model to show that CO/CH$_4$ radiative convection can significantly impact the pressure/temperature (PT) profile of a brown dwarf atmosphere. we also use idealized two-dimensional (2D) hydrodynamic simulations in the Ledoux regime to illustrate how the temperature gradient can be reduced when compositional source terms are active. 

%study the impact of chemical source terms on Ledoux convection. We extend the mixing length theory in order to include the convective thermal and compositional transport in 1D atmospheric codes. We use this new model to show that CO/CH$_4$ radiative convection is sufficient to reduce the temperature gradient in brown dwarfs atmospheres similarly to moist convection in Earth troposphere.

\section{Linear theory for adiabatic and diabatic thermo-compositional convection}\label{sect:linear}

We perform a linear stability analysis in the Boussinesq regime with source terms in order to derive the instability criterion for convection.
We start with the Euler equations with gravity and compositional and energy source terms:

\begin{eqnarray}\label{eq:euler}
 \frac{\partial \rho}{\partial t} + \vec{\nabla}\left(\rho\vec{u}\right) &=& 0\cr
 \frac{\partial \rho \vec{u}}{\partial t} + \vec{\nabla}\left(\rho\vec{u}\otimes\vec{u} + 
 P\right) &=& \rho \vec{g} \cr
 \frac{\partial \rho X}{\partial t} + \vec{\nabla}\left(\rho X\vec{u}\right) &=& \rho 
 R(X,P,T) \cr
 \frac{\partial \rho\mathcal{E}}{\partial t} + \vec{\nabla}\left(\vec{u}\left(\rho
 \mathcal{E}+P\right)\right) &=& \rho c_p H(X,P,T)
 \end{eqnarray}
 
with the total energy $\mathcal{E}=e+u^2/2+\phi$, $e$ the internal energy, $\vec{u}$ the velocity, $\phi$ the gravitational potential, the equation of state (EOS) of an ideal gas $\rho e (\gamma-1) =P$, and the gravity $\vec{g}=-\vec{\nabla}\phi$. With the ideal gas law, the temperature and mean molecular weight are linked by $P = \rho k_b T/\mu(X)$. $R(X,P,T)$ is the source term in the advection/reaction equation of the mass mixing ratio $X$, and $\rho c_p H(X,P,T)$ the source term in the energy equation e.g. thermal diffusion, radiative transfer heating rate or latent heat pumping and release. Note that the system is in general not conservative in the presence of these source terms.
We use these equations to model two different physical systems for which the variables and the equation of state will have slightly different meaning:
 
 \begin{itemize}
 \item miscible perfect gases with X the mass mixing ratio of one of the gases (e.g. CO or CH$_4$ in a brown dwarf atmosphere). The equation of state is closed with $1/\mu(X)=X/\mu_1+(1-X)/\mu_2$ with $\mu_1$ and $\mu_2$ the mean-molecular weight of the two components of the mixture.
 \item immiscible liquid drops in the Earth atmosphere with X the mass mixing ratio of water vapour in the air. The equation of state is closed with $1/\mu(X)=1/\mu_a+X/\mu_\mathrm{H_2 O}$ (the liquid drops are neglected in the EOS) with $\mu_a$ the mean molecular weight of dry air and $\mu_\mathrm{H_2 O}$ the mean molecular weight of water vapour.
 \end{itemize}

Going towards the Boussinesq approximation and following \citet{1966PASJ...18..374K}, we first rewrite the system in terms of potential temperature and composition advection assuming that $c_p$ is constant:
\begin{eqnarray}
 \frac{\partial X}{\partial t} + \vec{u}\cdot \vec{\nabla}\left(X\right) &=& R(X,P,T) \cr
 \frac{\partial \log \theta}{\partial t} + \vec{u}\cdot\vec{\nabla}\left(\log \theta\right) &=& \frac{H(X,P,T)}{T}
\end{eqnarray}

 with the potential temperature $\theta=T(P_\mathrm{ref}/P)^{(\gamma-1)/\gamma}$, and $P_\mathrm{ref}$ a reference pressure. We linearize the system around a background state given by 
\begin{eqnarray}
  \frac{\partial P_0}{\partial z} &=& -\rho_0 g \cr
  R(X_0,P_0,T_0) &=& 0 \cr
  \vec{u_0} &=& 0 \cr
  H(X_0,P_0,T_0) &=& 0
\end{eqnarray}

 The linearization leads to the following system assuming incompressibility in the mass conservation:

\begin{eqnarray}
\vec{\nabla}\left(\vec{\delta u}\right) = 0 \cr
\frac{\partial\rho_0 \vec{\delta u}}{\partial t} + \vec{\nabla}\left(\delta P\right)
 - \delta\rho \vec{g} = 0\cr
 \frac{\partial \delta X}{\partial t} +
\vec{\delta u}\cdot\vec{\nabla}\left(X_0\right) =  R_X \delta X+ 
R_P \delta P+ R_T \delta T \cr
\frac{\partial \delta \theta/\theta_0}{\partial t} + \vec{\delta u}\cdot\vec{\nabla}\left(\log \theta_0\right)
 =  \frac{1}{T_0}\left(H_X \delta X+ H_P \delta P
 + H_T \delta T \right)\cr
\end{eqnarray}

with $R_{X,P,T}$ and $H_{X,P,T}$, the derivative of the compositional and thermal source terms with respect to composition, pressure or temperature respectively. 
We close the system with the equation of state:
\begin{equation}
\frac{\delta P}{P_0} = \frac{\delta \rho}{\rho_0} + \frac{\delta T}{T_0} - \frac{\partial 
\log \mu_0}{\partial X} \delta X 
\end{equation}
while we have by definition of the potential temperature
\begin{equation}
\frac{\delta \theta}{\theta_0} = \frac{\delta T}{T_0} - \frac{\gamma-1}{\gamma}\frac{\delta P}{P_0}
\end{equation}
In the Boussinesq regime, the pressure perturbations are only kept in the momentum equation to balance the gravitational force, we therefore eliminate the pressure perturbations following \citet{1966PASJ...18..374K} in the equation of state and in the potential temperature perturbation and we also assume $R_P=0$ and $H_P=0$ in the rest of the study. 
\begin{eqnarray}
0 &=& \frac{\delta \rho}{\rho_0} + \frac{\delta T}{T_0} - \frac{\partial 
\log \mu_0}{\partial X} \delta X  \cr
\frac{\delta \theta}{\theta_0} &=& \frac{\delta T}{T_0} 
\end{eqnarray}

We add all the rest of the derivation of the instability criteria in Appendix, i.e. going to Fourier space, getting dispersion relations and finding the criteria for instability. The analysis of the dispersion relation shows that the flow becomes unstable whenever one of the following two inequalities is satisfied: 

\begin{equation}
\nabla_T-\nabla_\mathrm{ad}-\nabla_\mu > 0
\end{equation}
with $1/h_p=-\partial \log P_0/\partial z$, $\nabla_T = -h_p\partial \log T_0/\partial z $, $\nabla_\mu = -h_p\partial \log \mu_0/\partial z $, and $\nabla_\mathrm{ad} = (\gamma-1)/\gamma$, or
 
\begin{equation}\label{eq:criterion_sources}
(\nabla_T-\nabla_\mathrm{ad})\omega_X^\prime-\nabla_\mu\omega_T^\prime <0 
\end{equation}

with
\begin{eqnarray}
\omega_X^\prime &=& R_X + T_0 R_T \frac{\partial \log \mu_0}{\partial X}\cr
\omega_T^\prime &=&  H_T+  \frac{1}{T_0}H_X\left(\frac{\partial \log \mu_0}{\partial X} 
\right)^{-1}
\end{eqnarray}

The first inequality corresponds to the Ledoux criterion (see Eq.~\ref{eq:ledoux} in the appendix for details). The second inequality is  linked to the presence of source terms and we define this criterion as the diabatic criterion. Because $R_X<0$ (see appendix), the diabatic criterion reduces to the Schwarzschild criterion $(\nabla_T-\nabla_\mathrm{ad}>0)$ if $H_X=0$ and $\partial \log \mu_0/\partial X = 0$. Therefore there are three possibilities for the diabatic criterion not to trivially reduce to Schwarzschild criterion:
\begin{itemize}
\item case 1:  $H_X\ll H_T T_0 \partial \log \mu_0 /\partial X$, the criterion then reduces to $(\nabla_T-\nabla_\mathrm{ad})\omega_X^\prime-\nabla_\mu H_T <0 $ and the driving quantity for the instability is the mean-molecular-weight gradient
\item case 2: $H_X\gg H_T T_0 \partial \log \mu_0 /\partial X$, the criterion then reduces to $(\nabla_T-\nabla_\mathrm{ad})R_X+ H_X(\partial X_0/\partial z)/(T_0 h_p)<0$ and the driving mechanism is the dependance of the energy source term on composition
\item case 3: the general case with $H_X\sim H_T T_0 \partial \log \mu_0 /\partial X$ and the dependance of energy source term on composition and the mean-molecular-weight gradient are equally important driving mechanisms.
\end{itemize}

\begin{figure}[t]
\centering
\includegraphics[width=\linewidth]{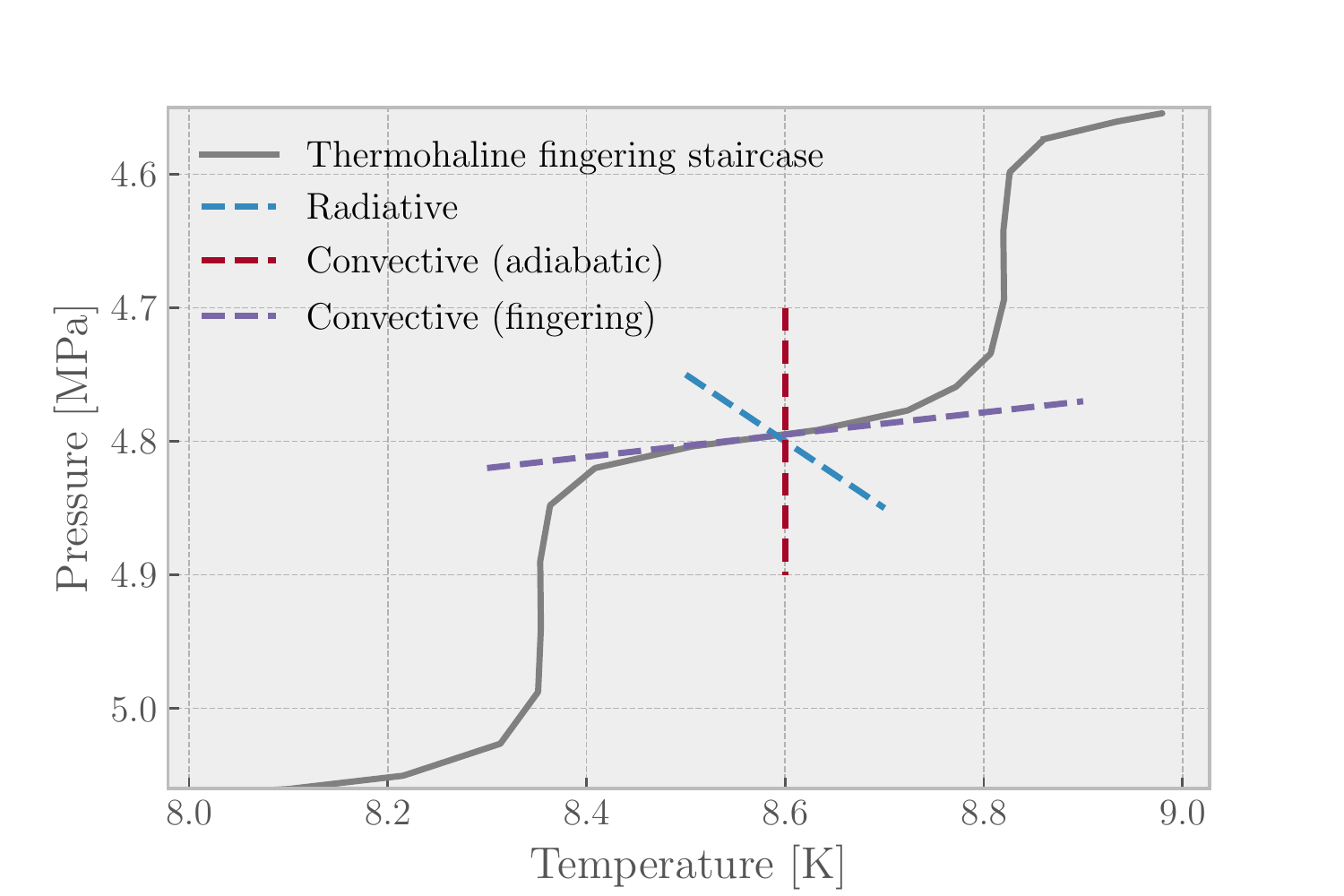}
\caption{\label{fig:pt_ocean} Pressure/Temperature profiles of a thermohaline fingering staircase adapted 
from \citet{GREGG1988453}.}
\end{figure}

\subsection{Thermohaline and fingering convection}

For double-diffusive convection (thermal diffusion coefficient $\kappa_T$ and compositional diffusion coefficient $\kappa_\mu$), the source terms are given by

\begin{eqnarray}
R(X) &=& \kappa_\mu \Delta X \cr
H(T) &=& \kappa_T \Delta T
\end{eqnarray}

we get $R_T=H_X=0$, $H_T=-k^2 \kappa_T$, and 
$R_X = -k^2\kappa_\mu$ with $k$ the norm of the wavenumber of the perturbation (see 
appendix). The instability criterion thus becomes

\begin{equation}
(\nabla_T-\nabla_\mathrm{ad})\kappa_\mu-\nabla_\mu \kappa_T>0 
\end{equation}

which is the standard form of the instability criterion for thermohaline and 
fingering convection \citep{Stern_1960}. Since $H_X=0$ in that case, we are in the case 1 situation and the driving mechanism is the mean-molecular-weight gradient.

In Earth oceans, the adiabatic PT profile is close to isothermal because of the incompressibility of liquid water. When oceans have hot and salty water on top of cold and fresh water, they can develop thermohaline convection through the instability criterion and develop the formation of salt fingers. When a blob of hot and salty water sinks, it will thermalize with the surroundings through temperature diffusion but remains heavier because salt diffusion is slower, hence the blob continues to sink and is unstable.

These fingers are thought to be at the origin of thermohaline-fingering staircases in the oceans \citep{TURNER1967599}. These structures alternate well-mixed layers at constant entropy (isothermal) and constant salt concentration and steps with sharp gradients of temperature and concentration prone to the fingering instability. An example of such a temperature structure is shown in Fig.~\ref{fig:pt_ocean} with the alternance of isothermal layers and fingering-favorable steps. Note that thermohaline staircases can also exist with the alternance of isothermal layers and the radiative steps shown in Fig.~\ref{fig:pt_ocean}. This situation occurs in polar Earth oceans when cold/fresh water overlies warm/salty waters \citep[e.g.][]{Timmermans_2008}. In the thermohaline fingering staircase, the well-mixed layers are prone to overturning convection and convective plumes can also be observed transporting energy through the layer \citep[see Fig. 12 in][]{GREGG1988453}. 
The temperature gradient (as a function of pressure) in the fingering steps are negative (hot water on top of cold water), hence they are reduced compared to the adiabatic (isothermal) gradient. Figure 3 of \citet{Radko_2014} shows the evolution of the averaged temperature profile in the formation of these steps. Starting from a given temperature gradient, the simulation evolves towards a reduced temperature gradient as a function of pressure (increased temperature gradient as a function of altitude) in the fingering layer and not towards the adiabatic isothermal profile. Hence we can conclude that fingering convection can reduce the temperature gradient relative to pressure in Earth oceans.

\subsection{Moist convection}

Let us assume that the reactive terms for composition are fast and at an equilibrium given 
by $X = X_\mathrm{eq} (P,T)$. We can then relate the compositional vertical gradient to temperature and pressure gradients using:

\begin{equation}
  \frac{\partial X}{\partial z} = \frac{\partial X_\mathrm{eq}}{\partial \log T}\nabla_T\frac{\partial \log P}{\partial z} + \frac{\partial X_\mathrm{eq}}{\partial \log P}\frac{\partial \log P}{\partial z}
\end{equation}

We also assume that the reaction source term $R$ is condensation/evaporation of water and 
the thermal source term is the corresponding release or pumping of latent heat $L$
\begin{equation}\label{eq:H=RL}
H = -\frac{R L}{c_p}
\end{equation}

and we neglect the temperature dependance of the latent heat. We can then derive from the criterion in Eq.~\ref{eq:criterion_sources} the standard form of the moist convective criterion used in Earth atmospheric physics  \citep[see][]{Stevens_2005}

\begin{equation}
\nabla_T -\nabla_\mathrm{ad}\frac{1+ \frac{X_\mathrm{eq}L}{R_d T_0}}
{1 +\frac{X_\mathrm{eq}L^2}{c_pR_v T_0^2}} >0 
\end{equation}

with $R_v$ and $R_d$ the vapour and dry air gas constant. We add the details of this derivation in appendix.

\begin{figure}[t]
\centering
\includegraphics[width=\linewidth]{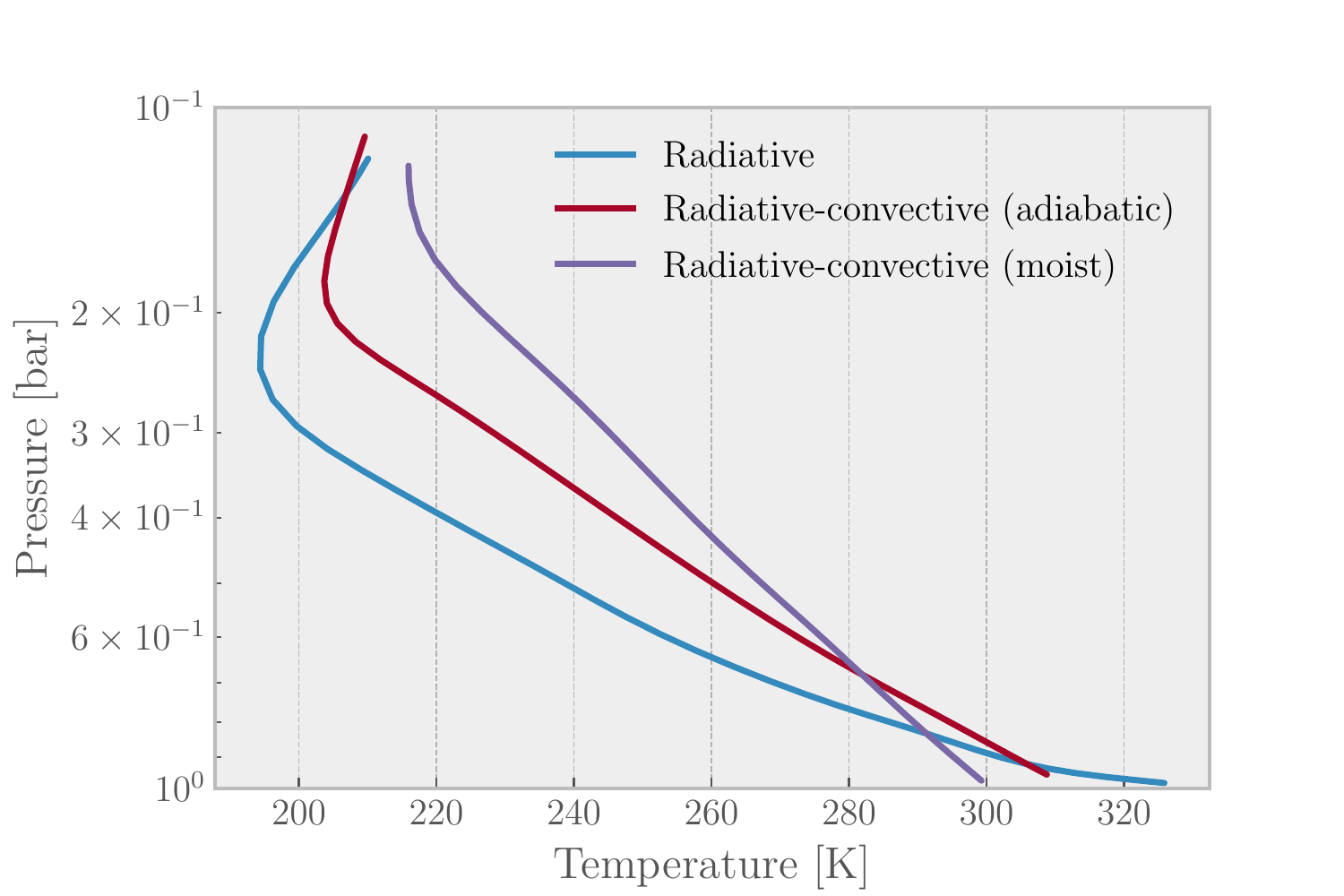}
\caption{\label{fig:pt_earth} Pressure/Temperature profiles for Earth atmosphere, adapted 
from \citet{Manabe_1964} with different convective adjustments.}
\end{figure}

Moist convection has been introduced and experimentally studied during the 50's and 60's. A historical perspective of moist convective studies can be found in Fig.~4 of \citet{YANO20141}. By taking into account the energy transported by water vapor through the potential latent heat release, it has been shown that the moist saturated PT profile
has a reduced temperature gradient (the lapse rate) compared to dry adiabatic convection. In Fig.~\ref{fig:pt_earth}, we show different convective adjustments for Earth atmosphere, adapted from \citet{Manabe_1964}. Dry adiabatic convection has a lapse rate of 10 K~km$^{-1}$ while moist convection leads to a reduced lapse rate, closer to the observed one in Earth troposphere (6.5 K~km$^{-1}$).

However, we insist on the fact that this pressure/temperature (PT) profile at the instability limit should not be called a moist adiabat because this creates confusion on the nature of the instability. As defined in \citet{Stevens_2005}, this is a pseudo-adiabat with an effective adiabatic gradient. If we were considering the gas phase and the liquid phase for the system, there would be no energy source term and the system would be adiabatic and obey a strict energy conservation. But since we are interested in the temperature of the gas phase for atmospheric studies, the latent heat that can be pumped or released from the liquid phase acts as an external source term. Hence the gas phase alone is not adiabatic (also because the liquid phase can decouple from the gas phase through precipitation).

An intuitive way to describe moist convection is as follows: when a parcel is rising, it cools by expansion thus triggers water condensation which releases latent heat and heat the atmosphere. This description fits in case 2 described in Sect.~\ref{sect:linear} which means that the driving mechanism is the dependance of the energy source term on composition.

\subsection{CO/CH$_4$ radiative convection}\label{sect:coch4}

%We can think of getting an analog of moist convection with CO/CH$_4$
%chemistry instead of water condensation/evaporation. This idea quickly fails on the fact 
%that the energy exchange in the CO/CH$_4$ reaction in an hydrogen dominated atmosphere 
%is orders of magnitude less than condensation/evaporation of water in Earth atmosphere. 
The key to understand CO/CH$_4$ radiative convection is to realize that moist convection is actually a particular case in our previous analysis for which the energy exchange is directly linked to the compositional reaction by Eq.~\ref{eq:H=RL}, with the energy source term being directly proportional to the compositional term. In the diabatic criterion in Eq.~\ref{eq:criterion_sources}, we do not need to assume such a direct link in the physics of the compositional source term and the thermal source term. We can thus derive an equivalent of the moist convective criterion with the two dominant processes at stake in brown dwarf atmospheres: CO/CH$_4$ chemistry and radiative transfer. A direct link between the two is not necessary since composition and temperature interact with each other either through $H_X \neq 0$, or through the equation of state with $\partial \log \mu_0/\partial X \neq 0$. When a parcel is rising, the temperature is changing because of expansion and radiative transfer, we then have two possibilities for CO/CH$_4$ radiative convection:
\begin{itemize}
\item  it triggers chemical exchanges between CH$_4$+H$_2$O and CO+H$_2$O, with opacity differences in the two states that can lead to heating and cooling in the atmosphere. The driving mechanism is the dependance of the energy source term on composition.
\item  similar to the thermohaline case, if the temperature adjustment is faster than the chemical timescale, the parcel will have more CO+H$_2$O than CH$_4$+H$_2$O compared to its local environment, hence a smaller mean-molecular-weight and continue to rise. In that case, the driving quantity for instability is the mean molecular weight gradient.
  %The entropy of the CO+H$_2$O dominated parcel is higher than the CH$_4$+H$_2$O dominated environment, hence when the parcel starts to be converted into CH$_4$+H$_2$O on a chemical timescale, its temperature will be higher than the surroundings and heat the atmosphere. The driving mechanism is the mean molecular weight difference and is similar to the illustration in Fig.~\ref{fig:schematics}.
\end{itemize}

In the general case, for CO/CH$_4$ exchange and radiative transfer
 \begin{eqnarray}\label{eq:sources}
R &=& -\frac{X- X_\mathrm{eq}}{\tau_\mathrm{chem}}\cr
H &=& \frac{4\pi \kappa}{c_p} (J-\sigma T^4/\pi)
\end{eqnarray}

with $\tau_\mathrm{chem}$ the timescale of the chemical reaction, $\kappa$ the absorption of the gas and $J$ the mean radiative intensity. Note that in the optically thick regime, when the radiative energy is small compared to the gas energy, the energy source terms can be written as thermal diffusion with a diffusion coefficient $\kappa_T = c/(3 \kappa \rho)$. For simplicity, we first assume that the opacity does not depend on composition, hence $H_X=0$. In that case the instability is only driven by the mean-molecular-weight gradient and:

\begin{eqnarray}
\omega^\prime_X &=& -\frac{1}{\tau_\mathrm{chem}} \cr
\omega^\prime_T &=& -\frac{1}{\tau_\mathrm{rad}}
\end{eqnarray}

with $\tau_\mathrm{rad}=c_p/(16 \pi \kappa \sigma T^3)$ in the optically thin regime, or $\tau_\mathrm{rad}=1/(k^2 \kappa_T)$ in the optically thick regime.
 
 Assuming that the chemistry is close to equilibrium in the deep atmosphere of brown dwarfs (similar to moist convection at saturation), the criterion then becomes:

\begin{eqnarray}
\nabla_T-\nabla_\mathrm{ad}\frac{1+ \frac{1}{\nabla_\mathrm{ad}}\frac{\partial\log\mu_0}
{\partial X}\frac{\partial X_\mathrm{eq}}{\partial \log P}\frac{\tau_\mathrm{chem}}
{\tau_\mathrm{rad}}}{1- \frac{\partial\log\mu_0}{\partial X}\frac{\partial X_\mathrm{eq}}
{\partial \log T}\frac{\tau_\mathrm{chem}}{\tau_\mathrm{rad}}} &>&0 
\end{eqnarray}

which has the same form as the moist convective instability criterion, the demonstration of this expression is the same as the demonstration for moist convection in appendix\footnote{It can also be shown that the same inequality can be derived when $H_X \neq 0$ and $\partial \log \mu_0/\partial X = 0$}. We show with Fig.~\ref{fig:pt_earth} that moist convection is known to reduce the temperature gradient in Earth troposphere and that fingering convection does the same in Earth oceans (see Fig.~\ref{fig:pt_ocean}). It becomes thus natural to expect CO/CH$_4$ radiative convection to behave in the same way since all these instabilities derive from the same criterion. As shown in \citet{tremblin:2015aa,tremblin:2016aa,2017ApJ...850...46T}, such a temperature gradient reduction in the atmospheres of brown dwarf can explain very well the spectral reddening of field brown dwarfs and also the strong reddening in the spectrum of young brown dwarfs in moving groups. However, we need to go to the non-linear regime in order to check if the diabatic convective fluxes of CO/CH$_4$ radiative convection are indeed sufficiently significative to reduce the temperature gradient in the atmospheres of brown dwarfs.

%In addition to the parallel with moist convection, the analogy with two-phase convective 
%flows in the cooling systems of nuclear power plants is also shedding a new light on the 
%nature of the L/T transition. This sharp transition may be viewed as a giant analog of the 
%boiling crisis described in Fig.~\ref{fig:boiling_crisis}. When brown dwarfs are sufficiently
%cold to form an "insulating" layer of CH$_4$, the CO/CH$_4$ radiative convection will start 
%to become inefficient and the atmosphere will start to adjust back onto the adiabatic 
%PT profile. Similarly to the sharp temperature increase of the heating plate,
%this would lead to a sudden increase of temperature in the deep atmosphere of brown dwarfs 
%explaining the J-band brightening in their spectrum at the L/T transition.

\section{Non-linear regime and mixing length theory}\label{sect:non-linear}
 
 %\subsection{2D hydrodynamic simulations}\label{sect:simu}

%\begin{figure}[t]
%\centering
%\includegraphics[width=\linewidth]{nukiyama_conv.pdf}
%\caption{\label{fig:nukiyama_conv} Adiabatic and diabatic convective fluxes 
%from the mixing length theory as
%an explanation of Nukiyama-type curves showing a bifurcation of the surface temperature as a function
%of imposed flux.}
%\end{figure}

\subsection{Conserved quantities in the linear regime}\label{sect:conserved}

Before going to the mean field approach, we will define two new conserved quantities in the linear regime. Without source terms, adiabatic convection will tend to homogenize potential temperature and composition, however this is not the case in general with  diabatic energy and compositional transports.
Assuming we are in the linear regime, we can rewrite the evolution of the perturbations with
\begin{eqnarray}
\frac{\partial \delta X}{\partial t} + \delta\vec{u}\cdot\vec{\nabla}X_0 &=& R_X \delta X + R_T \delta T\cr
\frac{\partial \delta \log \theta}{\partial t} +  \delta\vec{u}\cdot\vec{\nabla}(\log \theta_0) &=& \frac{1}{T_0}\left(H_T \delta T + H_X\delta X\right)
\end{eqnarray}

We make the ansatz that the density perturbations are small in the equation of state, which is valid with a background marginally unstable i.e. close to neutral buoyancy:
\begin{eqnarray}\label{eq:eos_link}
\frac{\delta\rho}{\rho_0} = \frac{\partial \log \mu_0}{\partial X} \delta X- \frac{\delta T}{T_0} \sim 0
\end{eqnarray}

This ansatz can be verified a posteriori in the numerical simulations performed in Sect.~\ref{sect:models}. We can then rewrite the evolution of the perturbations in a pure transport form:
\begin{eqnarray}
\frac{\partial \delta X^\prime}{\partial t} + \delta\vec{u}\cdot (\vec{\nabla}X^\prime_0 )&=& 0 \cr
\frac{\partial \delta \log \theta^\prime}{\partial t} +  \delta\vec{u}\cdot\vec{\nabla}(\log \theta_0^\prime)  &=& 0
\end{eqnarray}

with
\begin{eqnarray}\label{eq:primed}
X^\prime &=& X - \log \theta \left(\frac{\partial \log \mu_0}{\partial X}\right)^{-1}
\frac{\omega^\prime_X}{\omega^\prime_T}\cr
\log \theta^\prime &=& \log \theta - X \frac{\partial \log \mu_0}{\partial X}
\frac{\omega^\prime_T}{\omega^\prime_X}
\end{eqnarray}

This is the generalization of the moist potential temperature, in the case of moist 
convection it reduces to:
\begin{equation}
\log \theta^\prime = \log \theta - X \frac{L}{c_p T}
\end{equation}

This new potential temperature includes a term that depends on composition, that takes into account the potential release of energy from a compositional change through the energy and chemical source terms in the small-scale fluctuations. When the energy source term is proportional to the compositional source term (e.g. for moist convection), we do not need to linearize the system and we can define these conserved quantities directly in the non-linear regime. For arbitrary source terms, these quantities are only strictly conserved during the linear regime.

 \subsection{Extension of the mixing length theory}\label{sect:mlt}
 
 In sect.~\ref{sect:coch4}, we have assumed that the chemistry is at equilibrium in order
 to derive a criterion for CO/CH$_4$ radiative convection. This is valid for the deep atmospheres of brown dwarfs. It is well known, however, that CO/CH$_4$ chemistry is not at equilibrium in the upper part of the atmosphere. We therefore need to take into account out-of-equilibrium chemistry to compute the convective fluxes and we can use mixing length theory derived from a mean field approach to do that.
 
 We start from the equations of potential temperature and composition transport:
\begin{eqnarray}
\frac{\partial X}{\partial t} + \vec{u}\cdot\vec{\nabla}X &=& R(X,T)\cr
\frac{\partial \log \theta}{\partial t} + \vec{u}\cdot\vec{\nabla}(\log \theta) &=& \frac{H(X,T)}{T}
\end{eqnarray}

and we decompose the fields into a background component ($\bar{X}$, $\bar{T}$, $\bar{\vec{u}}=0$) and a small scale fluctuation part ($\delta X$, $\delta T$, $\delta \vec{u}$). As in the Boussinesq approximation we neglect the pressure perturbation $\delta P =0$ and choose the reference pressure in the potential temperature $P_\mathrm{ref} = \bar{P}$. We do not assume the amplitude of the fluctuations to be small compared to the background.

We can then rewrite the evolution of the large scale system with
\begin{eqnarray}
\frac{\partial \bar{X}}{\partial t} + \frac{1}{\rho}\frac{\partial X_d}{\partial z} &=& R(\bar{X},\bar{T})\cr
\frac{\partial \bar{T}}{\partial t} +\frac{1}{\rho c_p}\frac{\partial F_d}{\partial z}   &=& H(\bar{X},\bar{T})
\end{eqnarray}

with 

\begin{eqnarray}\label{eq:meanfield}
  \frac{1}{\rho c_p}\frac{\partial F_d}{\partial z}  & =& \frac{\partial \delta T}{\partial t}+ \delta\vec{u}\cdot\vec{\nabla}(\log\theta)T - H(X,T)+H(\bar{X},\bar{T})\cr
   \frac{1}{\rho}\frac{\partial X_d}{\partial z} &=& \frac{\partial \delta X}{\partial t}+ \delta\vec{u}\cdot\vec{\nabla}X - R(X,T)+R(\bar{X},\bar{T})
\end{eqnarray}

$F_d$ and $X_d$ are the turbulent convective fluxes that will contain the non-linear interactions between the background and the fluctuations and between the fluctuations themselves. The term $ H(X,T)-H(\bar{X},\bar{T})$ can be Taylor-expanded and contains all the non-linear interactions through the source terms represented by $H_{X^k,T^l}(\delta X)^k/k! (\delta T)^l/l!$ with $H_{X^k,T^l}$ being the k$^\mathrm{th}$ and l$^\mathrm{th}$ derivative with respect to composition and temperature. We then need to find a closure relation expressing $X_d$ and $F_d$ as a function of the background variables. It can be obtained through numerical simulations, by taking a space and time average of the fluxes defined with Eq.~\ref{eq:meanfield}. We point out that in general, it will be needed to evaluate all the non-linear terms. To avoid this complicated procedure, we propose to use an argument similar to mixing length theory.

When the source terms are zero, i.e. in the adiabatic case, the closure relation that leads to the standard mixing length theory is:

\begin{eqnarray}
  X_\mathrm{ad} &=&  -h_p\rho w \frac{\partial \bar{X}}{\partial z}\cr
 F_\mathrm{ad} &=& -h_p\rho c_p \bar{T} w\frac{\partial \log \bar{\theta}}{\partial z}
\end{eqnarray}

with $w = \omega l_\mathrm{conv}$, $ l_\mathrm{conv}$ being the mixing length and $\omega$ the growth rate of the convective instability. These relations correspond also to the flux-gradient laws used for thermohaline convection \citep[e.g.][]{Radko_2014}.

By using this closure relation, the convective fluxes will tend to homogenize the composition and the potential temperature, and gives the standard form of the convective flux for adiabatic convection.

\begin{equation}
F_\mathrm{ad} = \rho c_p w \bar{T} (\nabla_T-\nabla_\mathrm{ad})
\end{equation}

We have to make a clear distinction now between the convective instability that constrains the velocity in the flux (Schwartzschild, Ledoux or diabatic) and the energy and composition that are transported, which can be impacted by the source terms (diabatic transport) or not (adiabatic transport). This will depend on the averaging timescale in the mean field approach: if the source terms typical timescale is shorter than the averaging timescale, they can impact the transport of energy and composition even in the Schwarzschild or Ledoux convective regimes. 

In the linear regime, diabatic energy and compositional transports do not tend to homogenize potential temperature and composition as shown in Sect.~\ref{sect:conserved}. This is also likely the case in the non-linear regime. For that reason, we extend the analysis presented there and we suggest to use in the mean field model the quantities we identified as conserved during the linear regime. Using this ansatz, the new closure relation for the small scale diabatic processes is then given by:

\begin{eqnarray}
  X_\mathrm{d} &=&  -h_p\rho w \frac{\partial \bar{X^\prime}}{\partial z}\cr
 F_\mathrm{d} &=& -h_p\rho c_p \bar{T} w\frac{\partial \log \bar{\theta^\prime}}{\partial z}
\end{eqnarray}

with the prime quantities defined in Eq.~\ref{eq:primed}. The diabatic convective transport is similar to the parametrization used for moist convection \citep{ARAKAWA2011263}. In the thermohaline context, using the standard potential temperature and composition in the flux-gradient laws  can lead to inconsistencies such as an ultraviolet catastroph when trying to predict the growth rates of the staircase modes at small scales \citep{Radko_2014}. The new parameterization proposed here could provide a way to solve these issues.

The convective diabatic fluxes can be written in a more usual form:
\begin{eqnarray}
  X_\mathrm{d} &=&  \rho w\left(\frac{\partial \log \mu_0}{\partial X}
\right)^{-1}(\nabla_\mu - (\nabla_T-\nabla_\mathrm{ad})\omega^\prime_X/\omega^\prime_T)
\cr
 F_\mathrm{d} &=& \rho c_p w \bar{T} (\nabla_T-\nabla_\mathrm{ad}-\nabla_\mu 
\omega^\prime_T/\omega^\prime_X)
\end{eqnarray}

For the convective velocities, in the Schwarzschild and Ledoux regimes we get (for $k_z<<k_{x,y}$):

\begin{equation}
w = l_\mathrm{conv} \sqrt{\frac{g}{h_p}(\nabla_T-\nabla_\mathrm{ad}-\nabla_\mu)}
\end{equation}

while for the diabatic convective instability, we can approximate the growth rate and the convective velocity with (see appendix for details):
\begin{equation}
w = l_\mathrm{conv}\frac{\nabla_\mu \omega^\prime_T -
(\nabla_T-\nabla_\mathrm{ad})\omega^\prime_X}{(H_T R_X -H_X R_T)h_p/g-(\nabla_T-\nabla_\mathrm{ad}-\nabla_\mu)}
\end{equation}

Ideally, the mixing length $l_\mathrm{conv}$ can be estimated from the typical scale at which the most unstable linear mode saturates non-linearly (e.g. the typical finger length in the fingering convective case). This estimation does not seem easy to obtain in the general case and we take  $l_\mathrm{conv}$ as an adjustable parameter as often done in the mixing length theory. The growth rate can also be evaluated numerically directly from the dispersion relation. Of course, this convective velocity is always positive and is set to zero as soon as the instability criterion is not met. The existence of the adiabatic and diabatic convective fluxes in addition to the non-convective radiative flux, can give rise to bifurcations between the three energy transport regimes:
 \begin{itemize}
\item spatial bifurcation with e.g. the formation of the thermohaline staircases \citep{TURNER1967599}.
\item temporal bifurcation which could explain the L/T transition during the cooling sequence of brown dwarfs \citep{tremblin:2016aa}. 
 \end{itemize}

\begin{figure}[t]
\centering
\includegraphics[width=\linewidth]{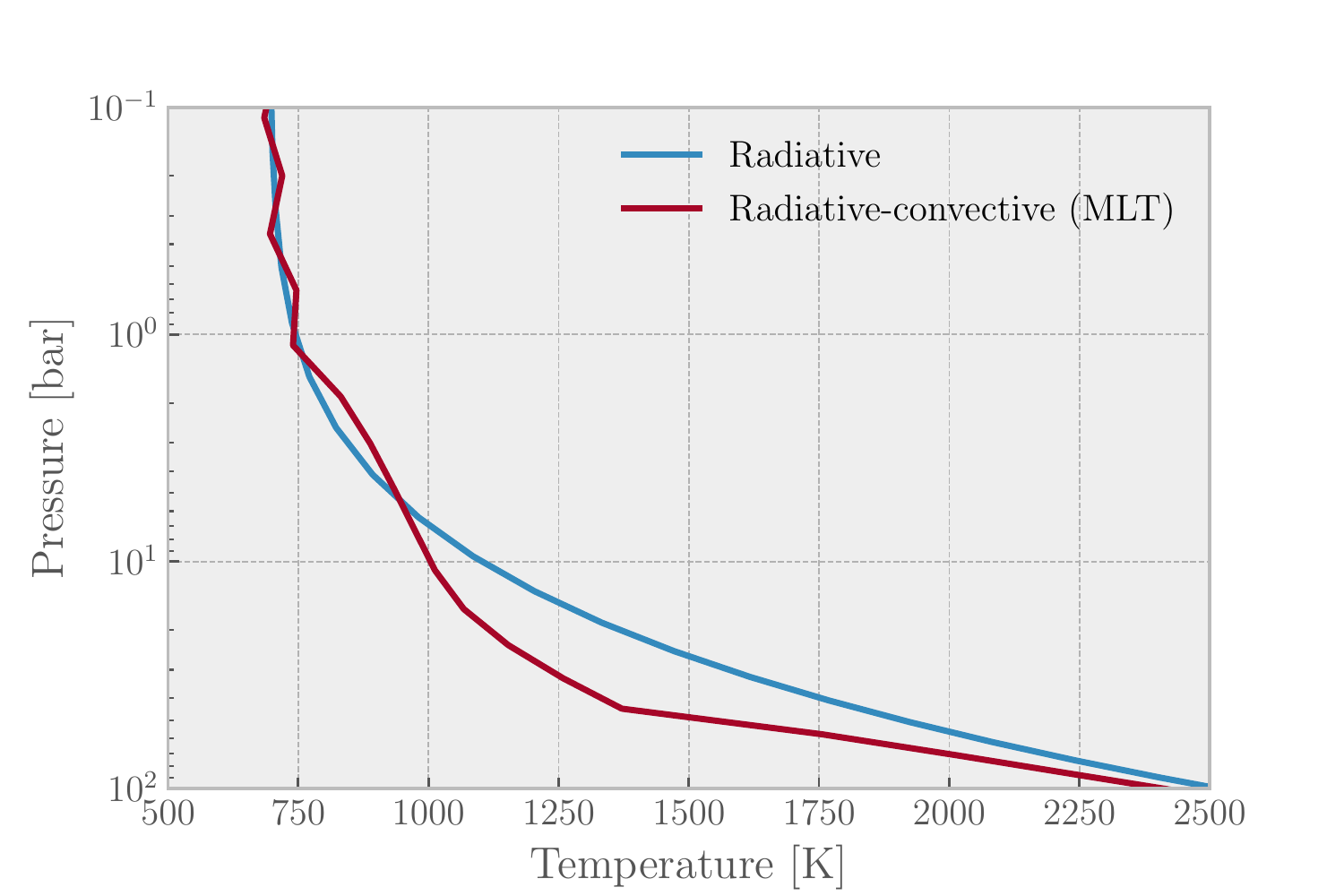}
\caption{\label{fig:pt_mlt} Radiative-convective equilibrium pressure/temperature profiles with 
and without the convective fluxes for CO/CH$_4$ radiative convection computed using mixing
length theory.}
\end{figure}

\begin{figure*}[t]
\centering
\includegraphics[width=0.49\linewidth]{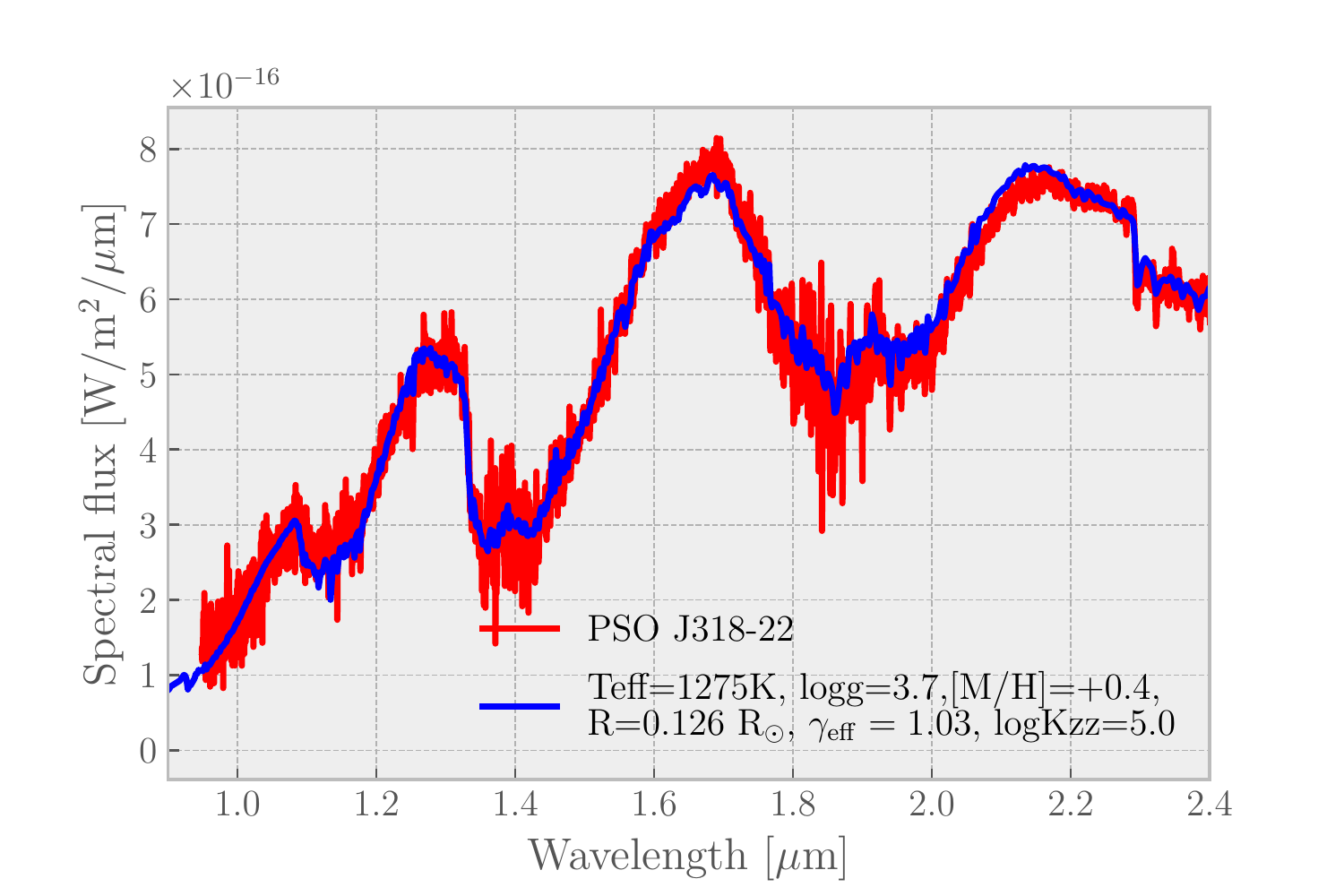}
\includegraphics[width=0.49\linewidth]{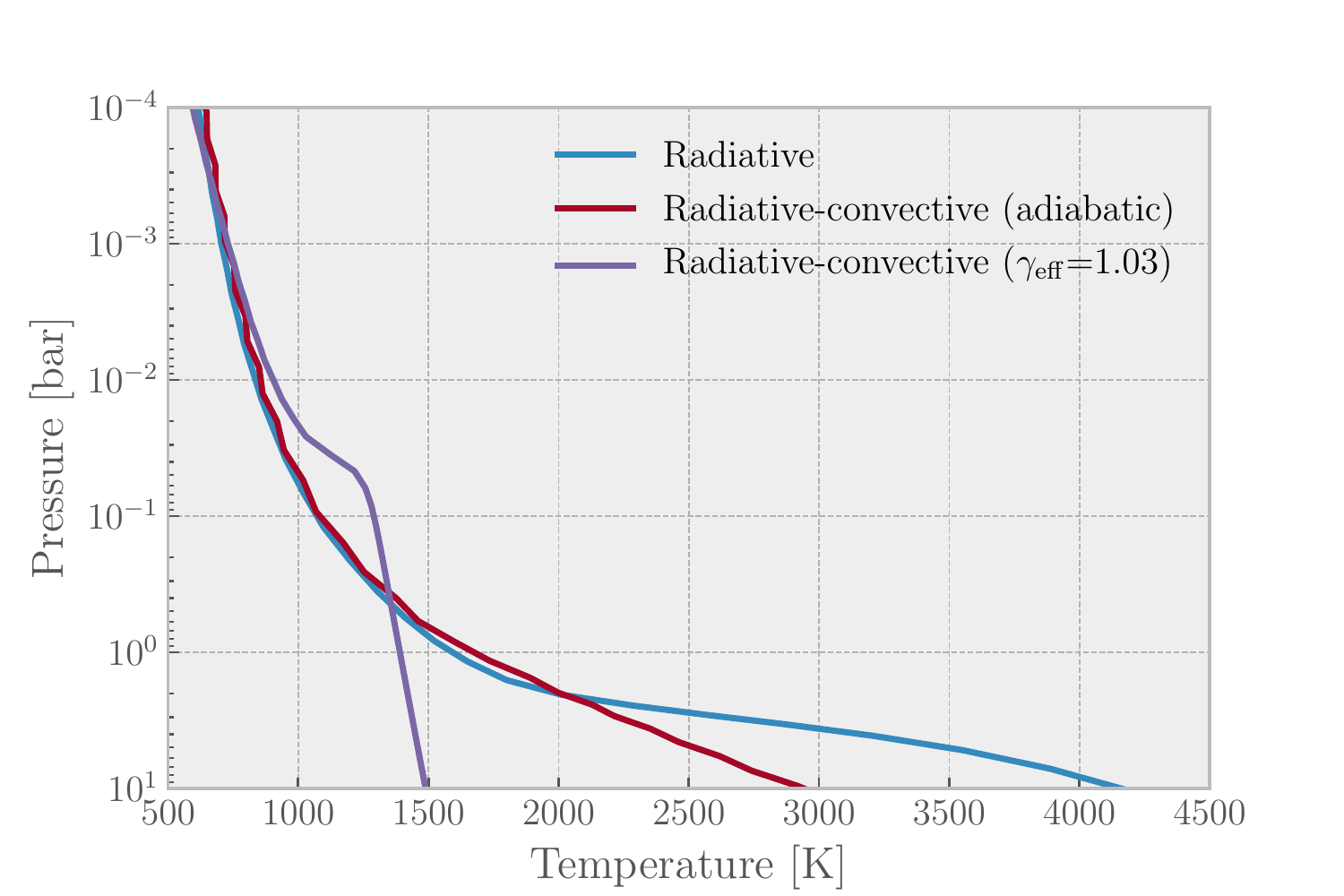}
\caption{\label{fig:pso} Left: Spectral model obtained with the \texttt{ATMO} code using 
a temperature gradient reduction in the atmosphere \citep{tremblin:2016aa} compared 
with  the GNIRS spectrum of PSO J318.5338-22.8603 (Liu et al. 2013). Right: Corresponding 
pressure/temperature profile with and without convective adjustments.}
\end{figure*}

\begin{figure*}[t]
\centering
\includegraphics[width=0.49\linewidth]{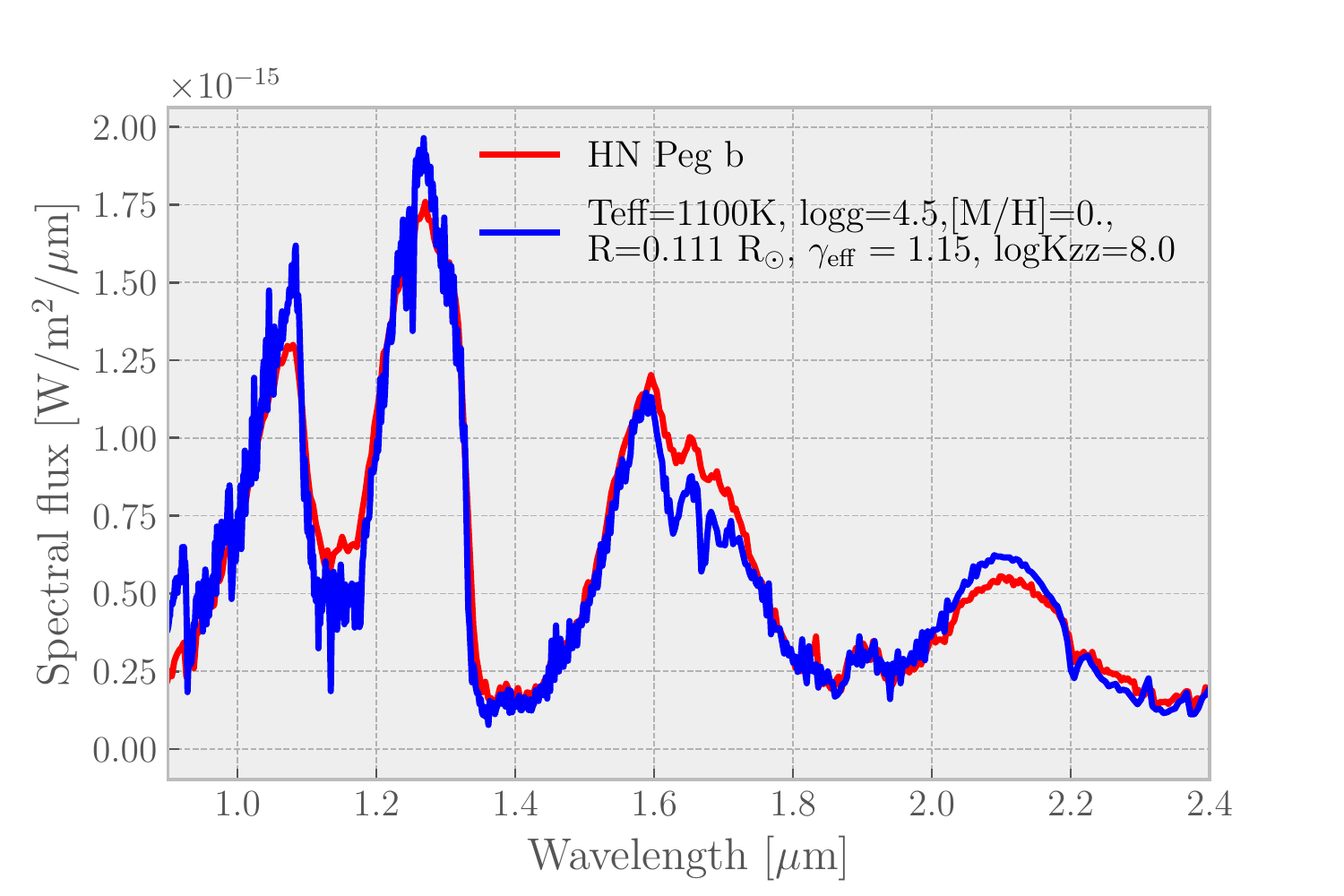}
\includegraphics[width=0.49\linewidth]{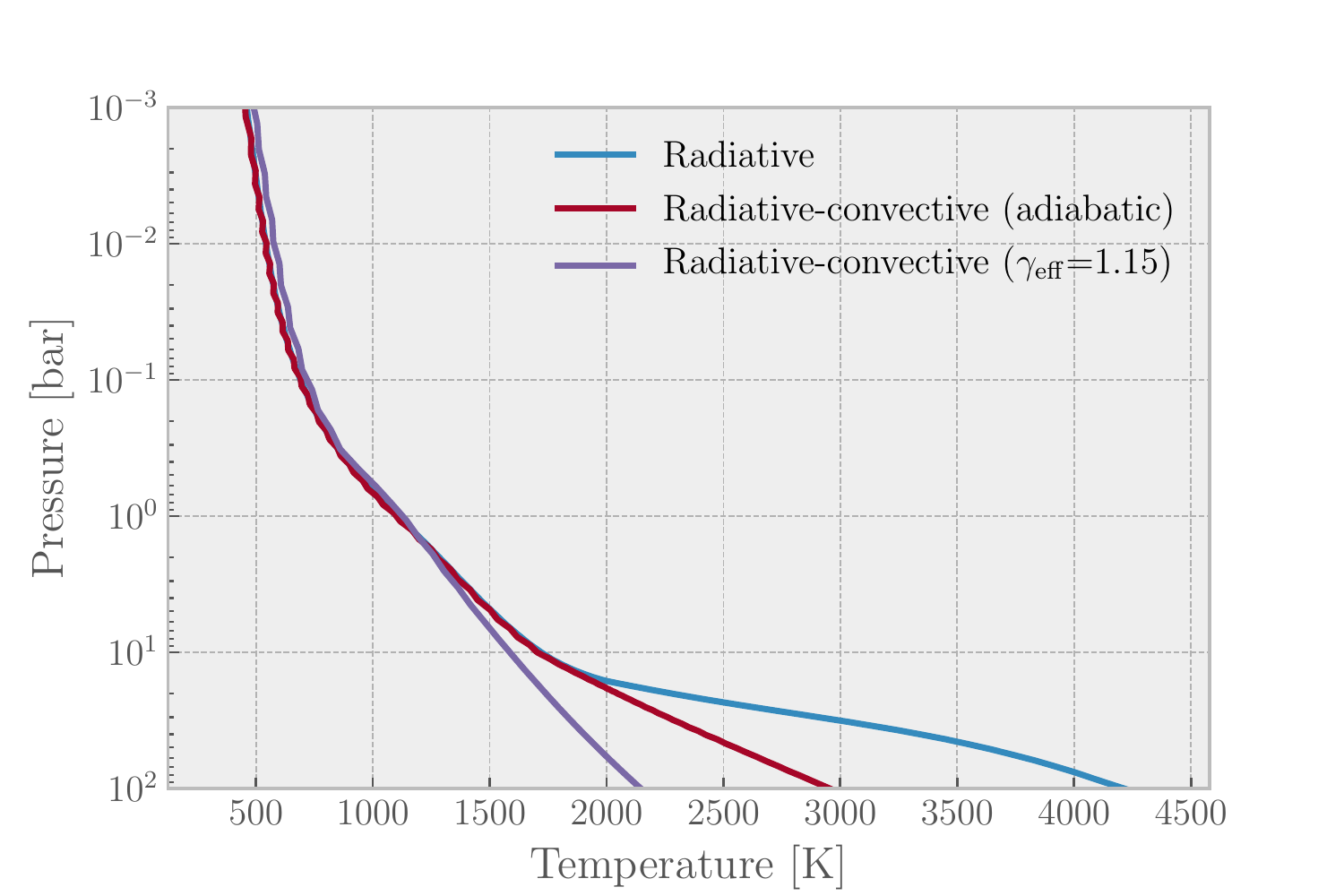}
\caption{\label{fig:hnpegb} Left: Spectral model obtained with the \texttt{ATMO} code using 
a temperature gradient reduction in the atmosphere \citep{tremblin:2016aa} compared 
with  the spectrum of HN Peg b. Right: Corresponding 
pressure/temperature profile with and without convective adjustments.}
\end{figure*}

 \section{CO/CH$_4$ radiative convection in brown dwarfs}\label{sect:models}

 \subsection{Application to brown dwarfs}

Using the small-scale parametrization of the convective transport we can now solve for the energy and compositional conservation equations in the whole atmosphere of a brown dwarf in steady state:

\begin{eqnarray}
\frac{\partial F_\mathrm{d}}{\partial z} &=& \rho c_p H \cr
\frac{\partial X_\mathrm{d}}{\partial z} &=& \rho R
\end{eqnarray}

Note that this system automatically includes out-of-equilibrium effects through the reactive  equation since the compositional convective flux can quench chemical source terms. We apply this mixing length model to the CO/CH$_4$ radiative convection with the source terms described by Eq.~\ref{eq:sources}.
Unfortunately, this system is quite stiff numerically because the factor $\omega^\prime_T/\omega^\prime_X=\tau_\mathrm{chem}/\tau_\mathrm{rad}$ can become very large when the chemistry starts to be slow. We have included the molecular diffusion timescale given by $l_\mathrm{conv}^2/\kappa_\mu$ to limit $\tau_\mathrm{chem}$ (with $\kappa_\mu$ the molecular diffusion coefficient). We have also implemented this mixing length parametrization in a toy model with grey radiative transfer and use a limiter approach to increase incrementally $\tau_\mathrm{chem}/\tau_\mathrm{rad}$ in the atmosphere. We show the result of this model in Fig.~\ref{fig:pt_mlt} on a grey atmosphere with an effective temperature at 825~K, gravity of 10$^4$ cm~s$^{-2}$, and a grey opacity profile of the form:
\begin{equation}
\kappa = \kappa_\mathrm{max} e^{-a z/z_\mathrm{max}}
\end{equation}
with $z_\mathrm{max}$ the thickness of the atmosphere in the model (1~\% of Jupiter radius), $z$ the local elevation from the bottom, $a$ is a dimensionless parameter fixed at 5, and $\kappa_\mathrm{max}$ 0.1 cm$^{2}$~g$^{-1}$ in order to get a temperature profile representative of a non-grey brown dwarf model. The pressure at the bottom of the atmosphere is fixed at 1~kbar. The temperature gradient of the purely radiative model corresponds to $\gamma_\mathrm{eff}$=1.4 with $\partial \log T/\partial \log P = (1-\gamma_\mathrm{eff}^{-1})$. For the mixing length model, we have used a mixing length at 0.016 times the local pressure scale height, and the limiter is set at $(\tau_\mathrm{chem}/\tau_\mathrm{rad})_\mathrm{max}$=10$^6$. The model with this value of the limiter is converged for this choice of mixing length, but we point out that with a higher mixing length it might be needed to increase the value of the limiter to get a converged model.

Figure~\ref{fig:pt_mlt} shows that we can indeed reduce the temperature gradient with CO/CH$_4$ radiative convection in a large part of the atmosphere, from 50 bars to approximately 1 bar, compared with the radiative or adiabatic profiles. The corresponding $\gamma_\mathrm{eff}$ value is going down to 1.1 in that model  (to be compared with the adiabatic exponent $\gamma\approx 1.4$) which is qualitatively what is needed to explain the reddening of brown dwarfs spectra as shown in Fig.~\ref{fig:pso} ($\gamma_\mathrm{eff}$=1.03 for PSO J318.5338-22.8603) and in Fig.~\ref{fig:hnpegb} ($\gamma_\mathrm{eff}$=1.15 for HN Peg b).

\subsection{Nature of the temperature gradient reduction}

The nature of the temperature gradient reduction is relatively clear in the case of moist convection. The direct release or pumping of latent heat clearly leads to such an effect. Even though, we can demonstrate with the generalized mixing length theory that the same phenomenom should arise for thermohaline and CO/CH$_4$ radiative convection, the explanation of this effect seems more subtle. In order to illustrate this mechanism, we present a series of idealized 2D stratified hydrodynamic simulations in the Ledoux unstable regime using the code \texttt{ARK}\footnote{\url{https://gitlab.erc-atmo.eu/erc-atmo/ark}, commit SHA 0248e6d1f2c97c39ad81be667795270f10efdd4c} developed in Padioleau et al. (submitted). The code is fully compressible and explicit, which means that we have a strong constraint on the size of the timestep that is limited by the speed of sound. It is therefore difficult to reach long timescales needed to explore the diabatic or fingering instability directly. Nevertheless, we can already perform simulations in the Ledoux unstable regime and show that compositional source terms can impact and reduce the temperature gradient even in that regime.
 
The system is composed of the Euler equations (Eq.~\ref{eq:euler}) with a gravitational and radiative source term and a reactive scalar field. The source terms are given by Eq.~\ref{eq:sources}. $X_\mathrm{eq}$ is a linear profile between 1 and 0 from the bottom to the top of the atmosphere and the mean-molecular weight is given by $1/\mu = X/\mu_1 +(1-X)/\mu_2$ and we will force a mean-molecular-weight gradient with $\mu_1<\mu_2$. When the chemical source term is included, we have used $\tau_\mathrm{chem}=20$~s. The radiative transfer is solved with a grey two-stream scheme adapted from the 1D/2D code \texttt{ATMO} \citep{tremblin:2016aa,drummond:2016aa} with the method from \citet{Trujillo_Bueno_1995}.
The incoming radiative flux at the bottom is set to a radiative temperature of $T_\mathrm{rad,zmin}=1100$~K and the downward radiative intensity at the top is set to zero. The opacity $\kappa$ is set to 5$\times$10$^{-4}$ cm$^2$~g$^{-1}$, i.e. $H_X=0$ and we explore the situation when the mean-molecular-weight gradient is the driving mechanism. The hydrodynamic is solved using a finite-volume "all regime" scheme well-suited for low-Mach and high-Mach flows \citep{Chalons_2016}, well-balanced for gravity \citep{chalons:hal-01297043}, (Padioleau et al. submitted). The well-balanced property of the solver allows the scheme to capture the hydrostatic balance $\partial P_0/\partial z = -\rho_0 g$ down to machine precision. With radiative transfer, we pre-compute the discretized PT profile satisfying both the hydrostatic balance and $H(X,P,T)=0$ for the discretized numerical scheme. By initializing the simulation on this profile, we are able to preserve the equilibrium state with vertical velocities lower than $10^{-5}$ cm~s$^{-1}$, even with an equilibrium state unstable to convection.  As a consequence, the quality of this numerical scheme allows us to study precisely the unstable convective mechanism and we need to explicitly introduce a velocity perturbation to trigger the instability
\begin{equation}
u_{0,z}(x,z)=A c_s \sin(m\pi x/x_\mathrm{max})e^{-(z-0.5z_\mathrm{max})^2/(W z_\mathrm{max})^2}
 \end{equation}
 with $c_s$ the local sound speed, $x_\mathrm{max}$, $z_\mathrm{max}$ the horizontal and 
 vertical extent of the box, respectively. For all the simulations in this paper, we have 
 used the same perturbation with $A=10^{-4}$, $W=0.25$, and $m=2$ in a box with 
 $x_\mathrm{max}=2.5\times 10^4$ cm, and $z_\mathrm{max}=5\times 10^4$ cm, and the pressure at the bottom of the atmosphere is fixed at $P_\mathrm{zmin} = 10$~bars. The total time of the simulations is 1600 s, i.e. 80 times the chemical timescale. All the simulations have reached a quasi-steady state at 800 s and we averaged the final profiles between 800 and 1600 s\footnote{The simulation outputs are available at \url{http://opendata.erc-atmo.eu}}. The resolution is fixed at 100$\times$50 cells and the scheme has a spatial and temporal discretization at first order.

 %No physical viscosity is included, and we have deliberately kept a low resolution to demonstrate that the instability mechanism is quite robust as it can develop with significant numerical diffusion.

 \begin{figure}[t]
\centering
\includegraphics[width=0.98\linewidth]{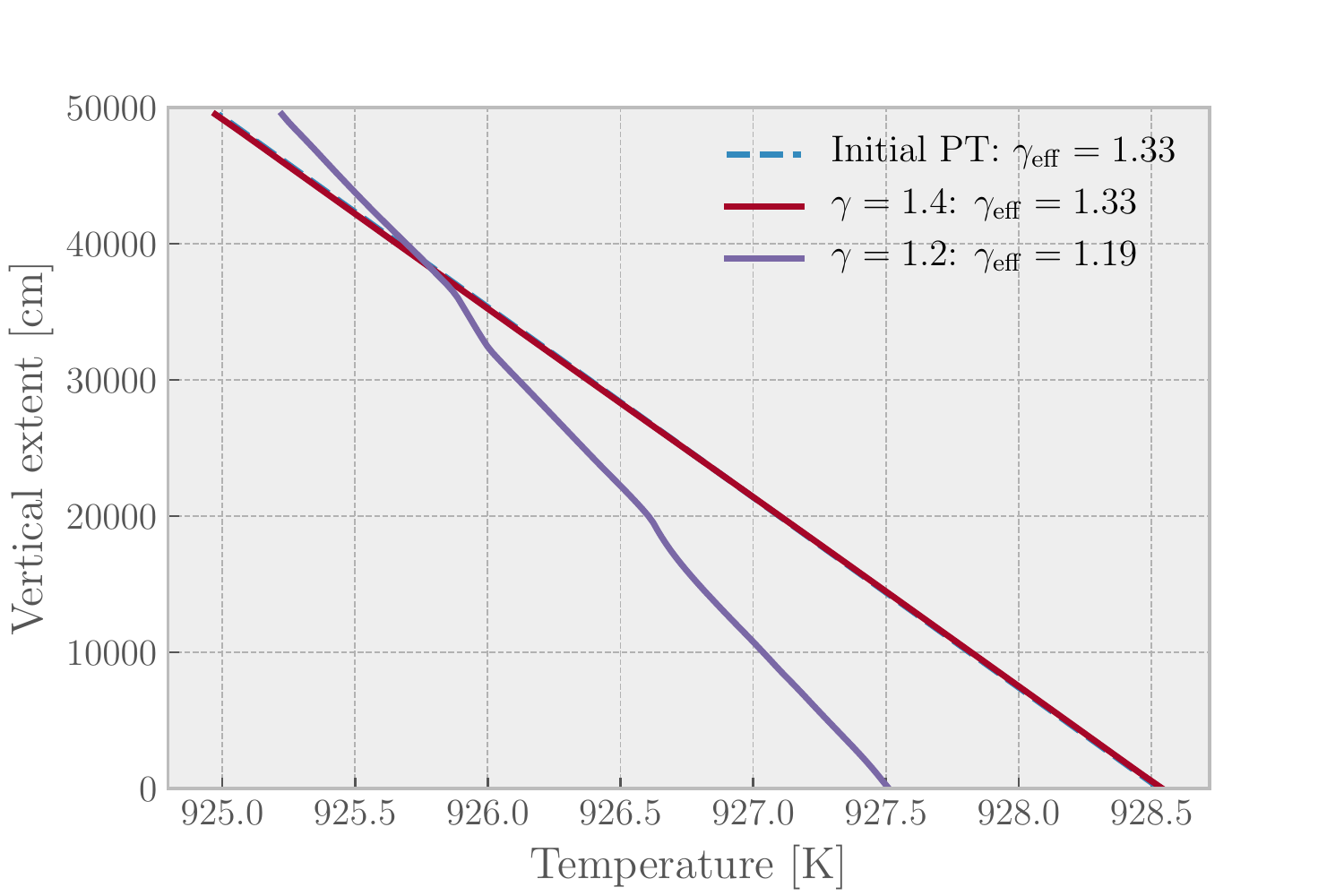}
\caption{\label{fig:temp2d_nochem} 1D averaged temperature profiles of 2D radiative convective simulations with no mean-molecular-weight gradient
and no chemical source term. The initial PT profile has an equivalent gamma index of $\gamma_\mathrm{eff}$=1.33.}
%When the adiabatic index of the gas is set to $\gamma=1.4$, no adiabatic convection can develop and the atmosphere
%stays at rest. When the adiabatic index of the gas is set to $\gamma=1.2$, the atmosphere is convectively unstable
%to adiabatic convection and convective motions reduce the temperature gradient down to $\gamma_\mathrm{eff}$=1.19.}
\end{figure}

\begin{figure}[t]
\centering
\includegraphics[width=0.98\linewidth]{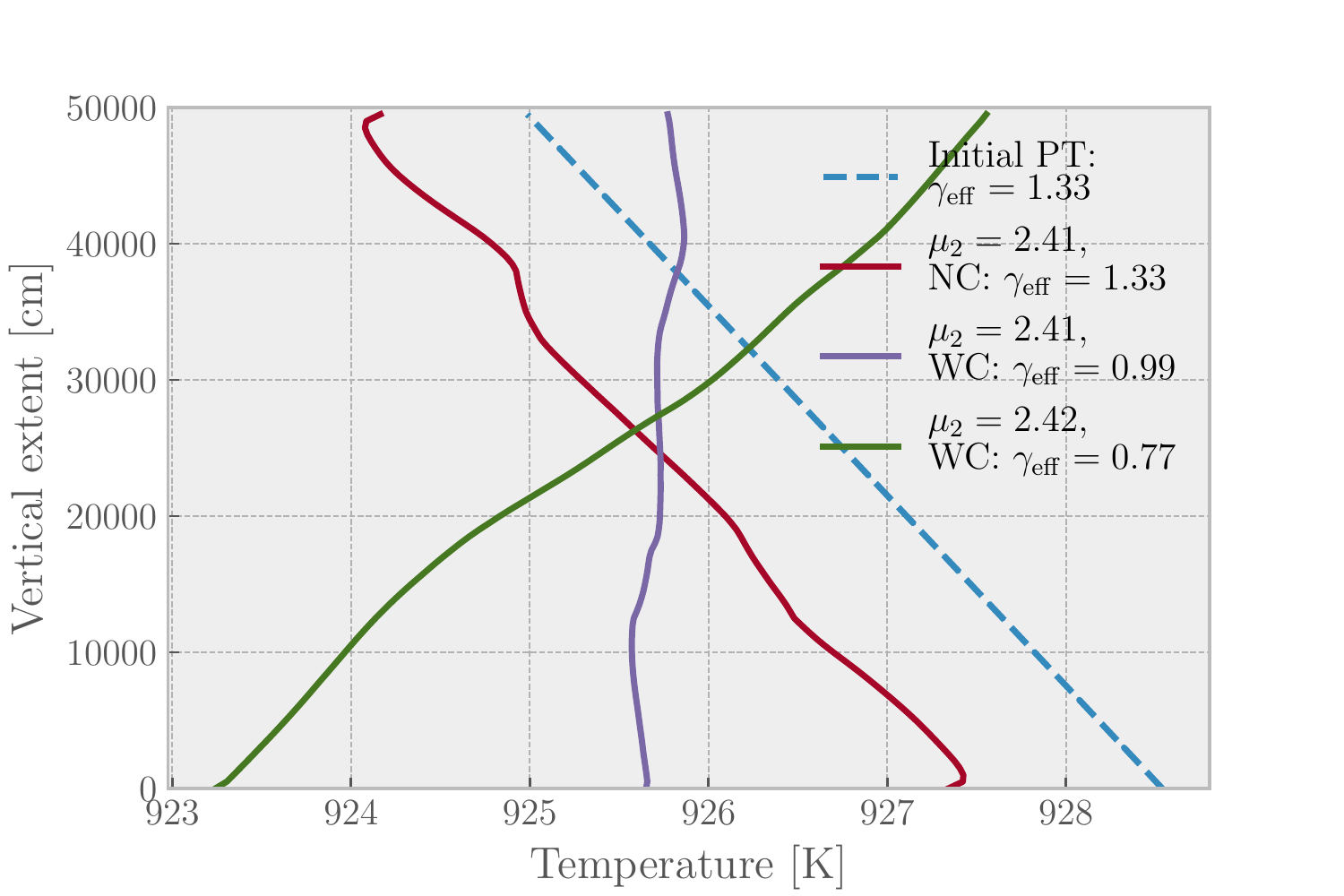}
\caption{\label{fig:temp2d_chem} 1D averaged temperature profiles of 2D radiative convective simulations with a linear  
mean-molecular-weight gradient between 2.41 and 2.42 at the top and 2.4 at the bottom of the box (Ledoux unstable).
The initial PT profile has an equivalent gamma index of $\gamma_\mathrm{eff}$=1.33. NC stands for No Chemistry and WC for With Chemistry.}
%Both simulations have an adiabatic index set to $\gamma=1.4$, stable to Schwarzschild convection, but both are Ledoux unstable.
%When the simulation has no chemical source term and the final averaged PT profile has a temperature gradient close to the initial gradient set by radiative forcing ($\gamma_\mathrm{eff}\approx1.33$). When the chemical source term is added and can trigger the energy exchange through chemical radiative convection, the final averaged PT profile has a temperature gradient reduction with $\gamma_\mathrm{eff}=0.99$.}
\end{figure}

%\begin{figure}[t]
%\centering
%\includegraphics[width=\linewidth]{time_vtot.pdf}
%\caption{\label{fig:time_vtot} Linear growth phase and saturation of the instability in the diabatic simulation with mean-molecular-weight gradient and chemical source term. The growth rate used for the red curve is computed with Eq.~\ref{eq:growth_rate} and corresponds to 0.025~s$^{-1}$.}
%\end{figure}
\begin{figure*}[t]
\centering
\includegraphics[width=0.75\linewidth]{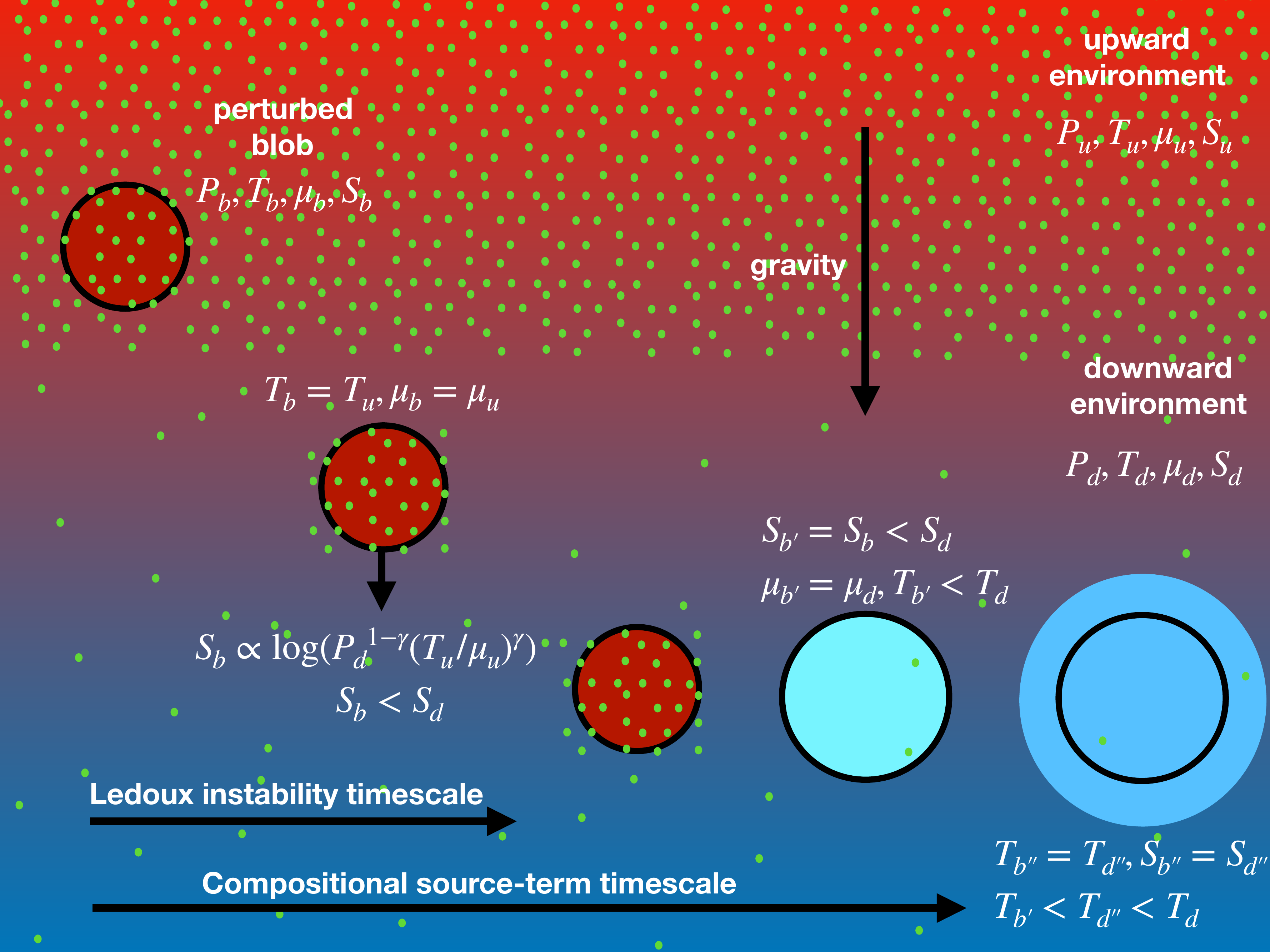}
\caption{\label{fig:schematics} Illustration of the Ledoux instability, including the effect of energy pumping by compositional change of a sinking blob in the surrounding environment (temporal evolution from left to right). This pumping is caused by the lower entropy of hot highly-concentrated materials compared to the entropy of the surrounding cold environment at low concentration. By symmetry a rising blob will release energy because of its higher entropy compared to the surrounding environment.}
\end{figure*}
  
We have performed first two simulations without mean-molecular-weight gradient 
($\mu_1=\mu_2=2.4$) and without reactive source term ($R(X,P,T)=0$). The initial PT profile of 
this setup corresponds to an equivalent gamma index of $\gamma_\mathrm{eff}=1.33$ (red curve in Fig.~\ref{fig:temp2d_chem}). We present the averaged temperature profile of the simulations in Fig.~\ref{fig:temp2d_nochem}. The first simulation is done with an adiabatic index of the gas $\gamma=1.4$, i.e. stable to Schwarzschild convection. The velocity perturbation is damped and the atmosphere stays at rest. The second simulation is done with an adiabatic index of the gas $\gamma=1.2$, unstable to Schwartzschild convection. Convective motions mix the atmosphere and the final PT profile has an equivalent gamma index of $\gamma_\mathrm{eff}=1.19$. These two simulations show that the code captures properly Schwarzschild convection similarly to Padioleau et al. (submitted) with Rayleigh-Benard convection. 
  
In the three other simulations in Fig.~\ref{fig:temp2d_chem}, the initial PT profile corresponds also to an equivalent gamma index of $\gamma_\mathrm{eff}=1.33$ and we have kept $\gamma=1.4$ so that the atmosphere is stable to Schwarzschild convection. We introduce in these simulations a 
mean-molecular-weight gradient with $\mu_1=2.4$ and $\mu_2=2.41$ or $\mu_2=2.42$ so that the atmosphere 
is unstable to the Ledoux criterion and we have verified that we get back the Ledoux growth rate in the linear regime. The first simulation is done with no reactive source term $R(X,P,T)=0$. In this situation, the final averaged PT profile has a temperature gradient similar to the initial one with $\gamma_\mathrm{eff}\approx 1.33$. The second simulation is done with the chemical source term and $\mu_2=2.41$. In that case, an exchange of energy can be performed through the compositional change induced by the chemistry which then
can lead to heating and cooling by the work of pressure adjustments due to expansion/contraction induced by the equation of state (increase/decrease of temperature at constant entropy). In this simulation, the final averaged PT profile has a significant reduced temperature gradient with an equivalent gamma index of  $\gamma_\mathrm{eff}=0.99$ (magenta curve in Fig.~\ref{fig:temp2d_chem}). In the third simulation, we have increased the mean-molecular-weight gradient with $\mu_2=2.42$. In that case the temperature gradient reduction is stronger with  $\gamma_\mathrm{eff}=0.77$ (green curve in Fig.~\ref{fig:temp2d_chem}) showing that the magnitude of the effect is a function of the mean-molecular-weight gradient.

A blob description helps explain why the temperature gradient can be decreased. Figure~\ref{fig:schematics} illustrates the behavior of a perturbed hot parcel with a high concentration of heavy element. Because the blob is Ledoux unstable, its entropy is lower than the surroundings. On a compositional timescale for the source term, this excess of composition will be dissipated in the non-linear regime. If we ignore first energy source terms (no source term on the entropy equation), the compositional change will happen at constant entropy, hence it will lead to a temperature of the blob smaller than the temperature of the surroundings, because of the expansion induced by the change in concentration (similar to haline expansion in the oceanic case) and the work of the pressure forces. If we re-introduce now thermal source terms, the blob will pump energy from the environment since it has a lower temperature, hence it will cool the deep cold atmosphere. That is the reason why the temperature gradient relative to pressure can be decreased even more in the presence of diabatic processes (even in the Ledoux regime). Of course the intermediate state ($T_{b^\prime},S_{b^\prime}$ in Fig.~\ref{fig:schematics}) in which we ignored energy source terms is a fictive state: the energy pumping by energy source terms will happen continuously during the compositional change phase if energy source terms are faster than compositional source terms. We suggest that this picture might also apply in the case of the formation of fingering steps in the thermohaline context.

%Note that the blob is also dissipating energy through thermal diffusion during the linear regime, the energy pumping during the non-linear regime will lead to a decrease of the global temperature gradient relative to pressure if it is bigger than the energy release during the linear phase. 

%The energy pumping of a sinking blob (or energy release by a rising blob) by activation of compositional source terms in the non-linear regime is not specific to fingering convection, we will show in Sect.~\ref{sect:simu} that we can reproduce this behavior with idealized 2D simulations in the Ledoux regime. 
  
 % We show in Fig.~\ref{fig:time_vtot} the time evolution of the mean of the norm of the velocity in the simulation. We can identify the linear phase of the instability with an exponential growth of the perturbation and the saturation at 200 seconds. The growth rate of the instability can be approximated by (see appendix):
%  
%  \begin{equation}\label{eq:growth_rate}
% \omega\approx \frac{g}{H_T R_X h_p}(\nabla_\mu \omega^\prime_T - 
%(\nabla_T-\nabla_\mathrm{ad})\omega^\prime_X)
%  \end{equation}
% For the diabatic simulation, the initial growth rate corresponds to 0.025~s$^{-1}$.
 
 \citet{Leconte_2018} has tried to argue that any type of convection should always lead
to an adiabatic PT profile. In general, an adiabatic PT profile requires at least an adiabatic energy equation, i.e. the energy source term should be zero or relatively small. This is not necessarily the case in the radiative part of an atmosphere or in the presence of latent heat release for Earth moist convection. We point out that this is also not the case for thermohaline convection \citep[see Fig. 3 in][]{Radko_2014}. The 2D simulations presented in this paper also clearly show that as soon as a compositional source term is included in the problem, which is absent in the analysis presented in \citet{Leconte_2018}, the temperature gradient can be reduced compared with an adiabatic one.

The 2D simulations we have performed are very idealized and in the Ledoux unstable regime. This is why we only use them to illustrate that a compostional source term can significantly impact the thermal profile of an atmosphere. The extension of the simulation box is 1 \% of a scale height and the mean-molecular-weight gradient is 10 times more than the gradient induced by the CO/CH$_4$ transition and the opacity is relatively small and we neglect so far the dependance of the energy source term on composition. Going to more realistic conditions will be the challenge for future works because it will require large computational resources to really explore the diabatic instability at stake in brown dwarf atmospheres.

\begin{figure}[t]
\centering
\includegraphics[width=\linewidth]{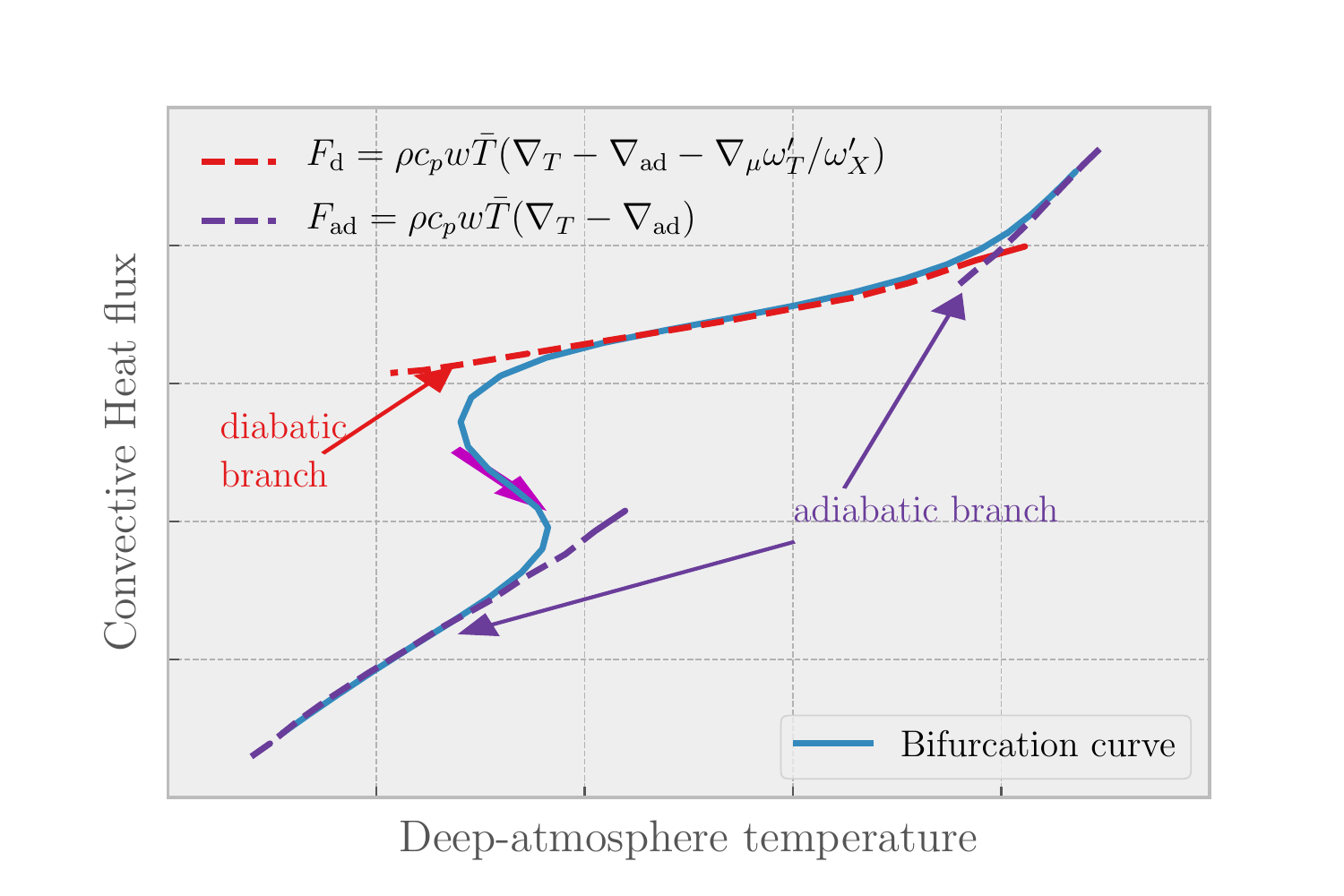}
\caption{\label{fig:nukiyama_bd_conv} Proposed bifurcation curve for the cooling sequence in brown-dwarf atmospheres showing a bifurcation of the deep-atmosphere temperature as a function of imposed flux. The bifurcation can explain the L/T transition and the warming of the atmosphere at the transition (magenta arrow) leading to the J-band brightening in the spectrum of brown dwarfs.}
\end{figure}

\begin{figure}[t]
\centering
\includegraphics[width=\linewidth]{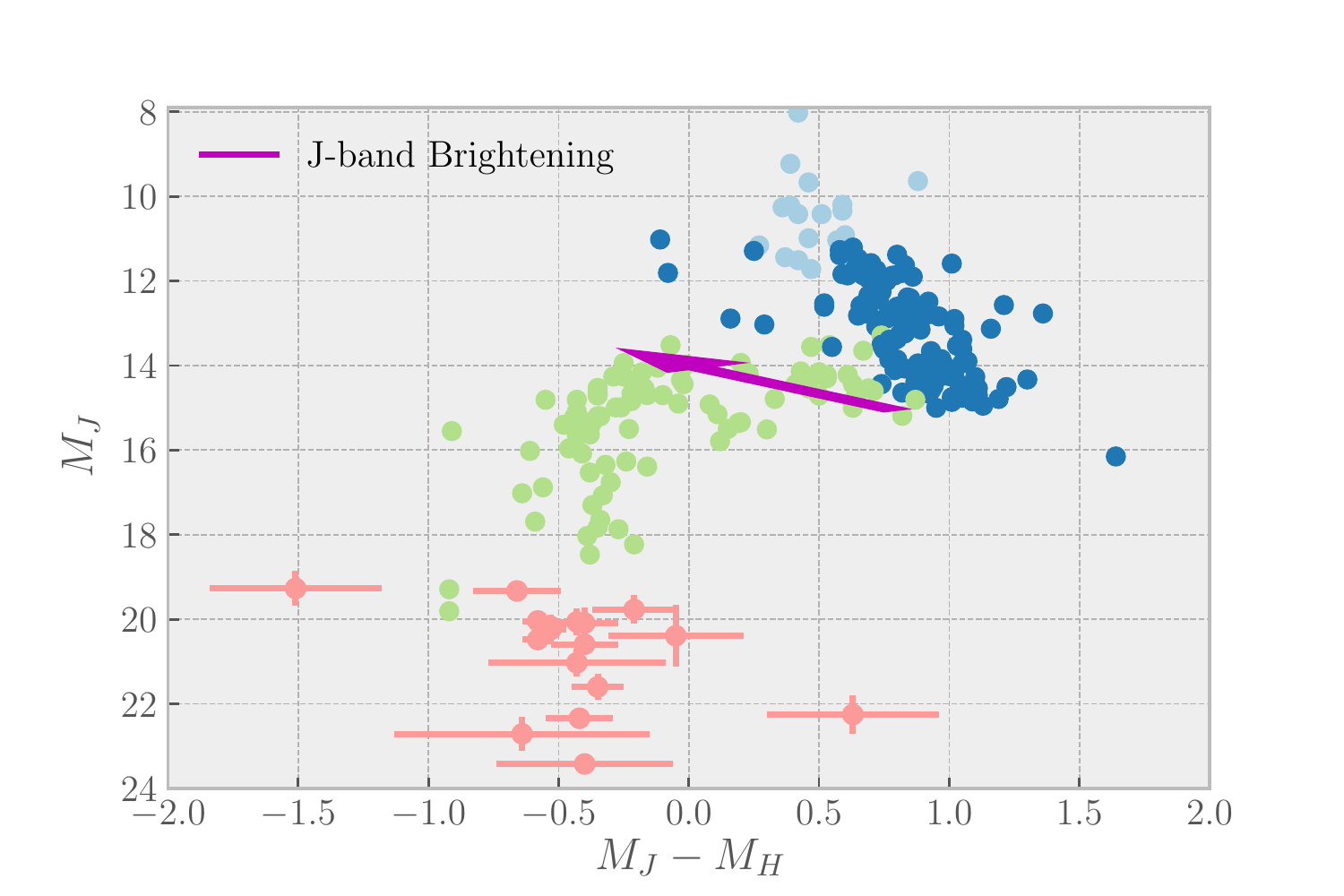}
\caption{\label{fig:col_mag_jh} Color magnitude diagram $M_J$ as a function of $(M_J-M_H)$. The magenta arrow indicates the J-band brightening in the color magnitude diagram. The photometry is extracted from \citet{Faherty:2012cy,Dupuy:2012bp,Dupuy:2013ks,Beichman:2014jr}. The different colors from top to bottom correspond to respectively, M, L, T, and Y dwarfs.}
\end{figure}

\section{Conclusions and discussion}

In this paper, we have developed a generalization of the theory of convection in order to incorporate any type of
compositional and energy source terms. We have shown that:
\begin{itemize}
\item Thermohaline convection in Earth oceans, fingering convection in stellar atmospheres, and moist convection in Earth atmosphere are all deriving from the same general instability diabatic criterion.
\item CO/CH$_4$ radiative convection in the atmospheres of brown dwarfs and extra-solar giant planets is also deriving from this diabatic instability and is therefore an analog to both Earth moist convection and thermohaline convection.
\item A generalization of mixing length theory is able to describe the energy and compositional convective transports and naturally includes out-of-equilibrium chemistry. Its application to CO/CH$_4$ radiative convection shows that we can qualitatively significantly reduce the temperature gradient in a large part of the atmospheres of brown dwarfs. The convective fluxes that we have derived can take into account the impact of the source terms on the transported quantities (diabatic energy and compositional transport) even in the case of Schwartzschild or Ledoux convection.
\item At small scales, 2D idealized hydrodynamic simulations of radiative convection in the Ledoux unstable regime is able to reduce the temperature gradient in the atmosphere compared with both the radiative or the adiabatic ones. This phenomenom can be understood as a pumping of energy of low-entropy sinking materials (or release of energy of high-entropy rising materials) induced by compositional changes. Therefore it demonstates that even in the Ledoux unstable regime, diabatic processes need to be taken into account in the convective transport (provided that the chemical timescale is fast enough). They also show that a convectively unstable system does not necessarily yield an adiabatic profile when source terms are taken into account in the process.
\end{itemize}

Based on the different form of the convective fluxes that we can derive using the generalized mixing length theory  (a given flux value can correspond to different temperature gradients), we can propose in Fig.~\ref{fig:nukiyama_bd_conv} the existence of a bifurcation for the cooling sequence of brown-dwarf atmospheres. Such a bifurcation could explain the J-band brightening observed in the spectra at the L/T transition since a warming of the atmosphere is produced at the transition when bifurcating from the diabatic convective branch to the adiabatic convective branch \citep[see also][]{tremblin:2016aa}. We show a color magnitude diagram in Fig.~\ref{fig:col_mag_jh} to illustrate the similarity between the flux/temperature curve and the observed cooling sequence in brown dwarf atmosphere. The comparison between the two figures highlight the correspondance between the J-band brightening and the warming of the deep atmosphere. Since it remains difficult to reproduce the J-band brightening with cloud models \citep[e.g.][]{charnay:2018aa}, if verified, the bifurcation between convective regimes would indicate that clouds are not responsible for the spectral reddening and the L/T transition even though they can be present in the atmospheres of brown dwarfs. Note that a similar bifurcation can take place with the N$_2$/NH$_3$ transition at the T/Y transition \citep{tremblin:2015aa}.

\begin{figure}[t]
\centering
\includegraphics[width=\linewidth]{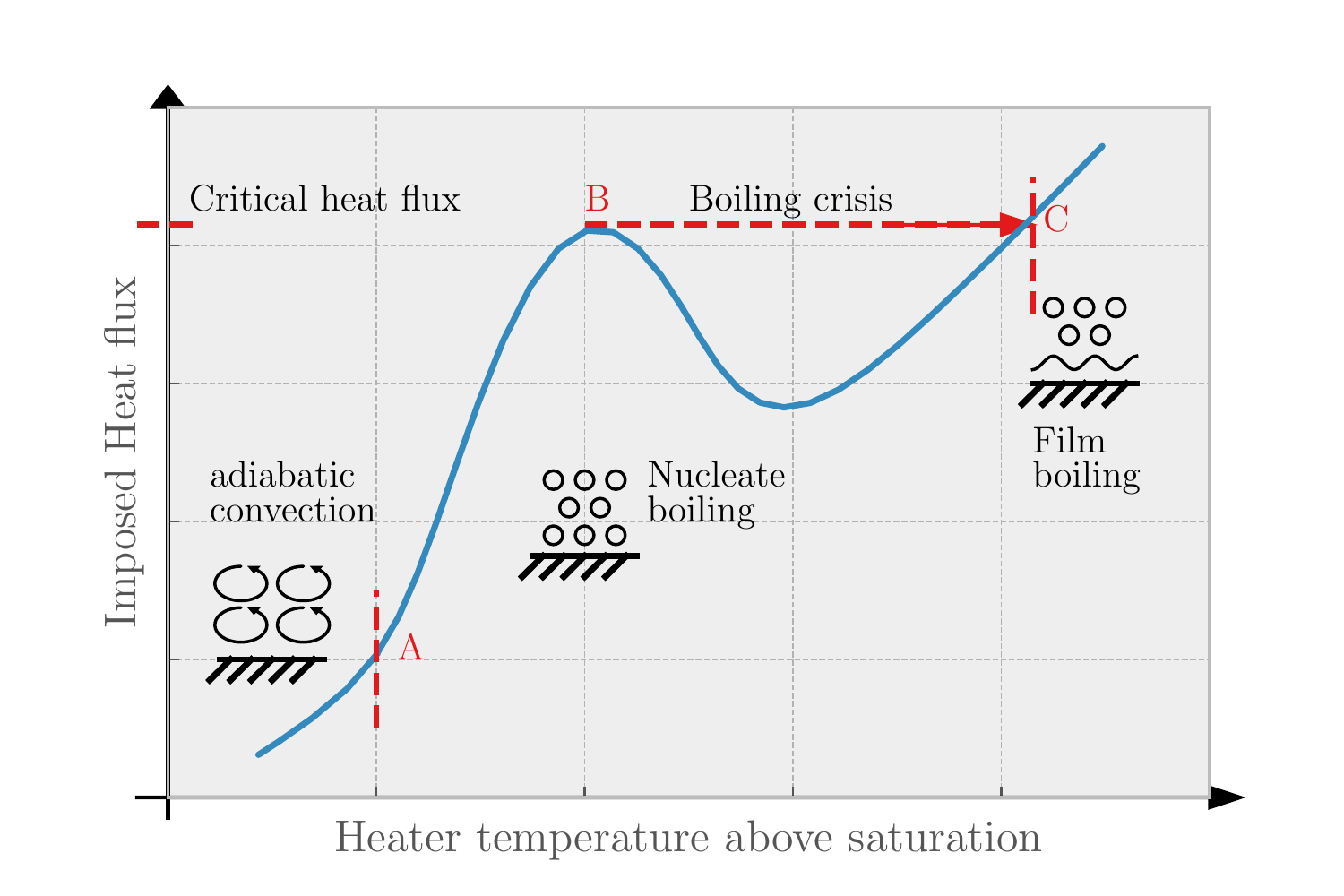}
\caption{\label{fig:boiling_crisis} Evolution of the heater temperature above the saturated 
fluid temperature as a fonction of heat flux in a boiling crisis experiment. This curve is
 known as the Nukiyama curve \citep{shiro:1934aa}.}
\end{figure}

Such a bifurcation seems to share analogies with the boiling crisis arising in two-phase convective flows. The transport of steam water in the water liquid phase is a problem of great importance for the cooling system of nuclear power plants \citep{Nikolayev_1999}. In Fig.~\ref{fig:boiling_crisis} we show the evolution of the temperature of a heating plate immersed in liquid water as a function of the imposed heat flux on the plate \citep{shiro:1934aa}. When the heating flux is sufficiently high (passing point A), evaporation and the creation of steam water bubble will arise, this phase is called nucleate boiling. The energy extraction by the convective flow is more efficient in that phase compared to adiabatic convection, the temperature of the heater is increasing slowly as a function of increasing heat flux. This regime is therefore quite attractive for the energy extraction from nuclear combustible. However, when the heat flux reaches the critical heat flux at point B, a catastrophic event happens known as the boiling crisis: a film of steam water is insulating the heating plate and its temperature is suddenly strongly rising up to point C. This bifurcation in the heating sequence is caused by the inefficient radiative energy transport in the vapour film, which seems to share some analogies with the formation of CH$_4$ during the cooling sequence in brown-dwarf atmospheres. The L/T transition could be seen as a giant cooling crisis, analog to the boiling crisis in two-phase convective flows.

We emphasize again that the Ledoux unstable simulations presented in this paper are idealized setups to study the pumping of energy in the presence of chemical reactions and are not directly applicable to brown dwarfs that are Ledoux stable but unstable to the diabatic criterion. Nonetheless these simulations demonstrate that the presence of chemical source terms can lead to a temperature gradient reduction, and as demonstrated by the 1D models with the extended mixing length theory, we can expect that the same phenomenon happens in the atmospheres of brown dwarfs that are subject to the diabatic CO/CH4 radiative instability. A lot of work remains to be done on CO/CH$_4$ radiative convection to quantify its effects on the temperature gradient and on comparisons with observations. We need large numerical simulations to reach the largest possible scale in the atmosphere with the diabatic instability and the incorporation of the generalized mixing length theory in 1D atmospheric codes with non-grey radiative transfer. We could then explore in details the role of the dependance of the energy source terms on composition ($H_X \neq$ 0). The convergence of the mixing length scheme might be difficult in 1D atmospheric codes because of the stiffness of the problem, but the generalized mixing length theory could also be of interest for other fields as it could be used as another parametrization of moist convection for Earth global circulation models and thermohaline convection for oceanic circulation.

%We have also suggested that the L/T transition can be interpreted as an equivalent
%of the boiling crisis, leading to the sudden warming up of the deep atmosphere similarly to the sudden warming of the heating plate in a pool boiling experiment. This effect may also have its equivalent for Earth climatic system. During deglaciation periods, Earth atmosphere is experiencing
%abrupt changes such as the rise of the level of oceans 
%and a sudden rapid global warming which can be interpreted in terms of bifurcation theory \citep[similarly to][for oceanic circulation]{Lenton_2012}.
%Seeing the ocean/atmosphere system as a global two-phase flow system, the presence of oceans can be
%interpreted as the equivalent of film boiling, with a layer of liquid phase in which convection 
%will be adiabatic and relatively inefficient compared to dry land that would be equivalent to nucleate boiling
%with moist convection. In that case the rise of oceans is equivalent to a transition from nucleate boiling
%to film boiling and the sudden abrupt rise of Earth surface temperature could be interpreted as a bifurcation
%from diabatic to adiabatic convection as in Fig.~\ref{fig:nukiyama_conv}. This transition in the 
%convective system could therefore explain the non-linearity of the temperature response to Milankovicth 
%forcing that is thought to be at the origin of glaciation/deglaciation cycles.
%The possible link between the boiling crisis, the L/T transition in brown dwarf atmosphere and deglaciation
%in Earth climatic system needs of course to be further explored. 

Chemical radiative convection might be a mechanism at work in many giant and Earth-like exoplanets, similarly to CO/CH$_4$ radiative convection in the atmospheres of brown dwarfs. The last interesting point is the possibility to take advantage of the wealth of good quality observational data on brown-dwarf atmospheres to use brown dwarfs as a laboratory to better understand the differences and similarities with Earth moist convection and thermohaline convection. Its evolution as a function of e.g., effective temperature and compositional change could provide valuable insights on some aspects of the physics at play in the climate of our own planet.

\acknowledgments
PT and HB acknowledges supports by the European Research Council under Grant Agreement ATMO 757858. This work is also partly supported by the ERC grant 787361-COBOM. BD acknowledges support from the STFC Consolidated Grant ST/R000395/1.

\appendix\label{appendix}

 The linearization of the Euler equations with gravity and compositional and energy source terms in the Boussinesq regime leads to the following system:

\begin{eqnarray}
\vec{\nabla}\left(\vec{\delta u}\right) &=& 0 \cr
\frac{\partial\rho_0 \vec{\delta u}}{\partial t} + \vec{\nabla}\left(\delta P\right)
 - \delta\rho \vec{g} &=& 0\cr
 \frac{\partial \delta X}{\partial t} +
\vec{\delta u}\cdot\vec{\nabla}\left(X_0\right) -  R_X \delta X -  R_T \delta T &=&0\cr
\frac{\partial \delta T}{\partial t} + T_0\vec{\delta u}\cdot\vec{\nabla}\left(\log \theta_0\right) - H_X \delta X -  H_T\delta T &=&0\cr
 \frac{\delta \rho}{\rho_0} + \frac{\delta T}{T_0} - \frac{\partial 
\log \mu_0}{\partial X} \delta X  &=&0
\end{eqnarray}

Then we assume the form $\exp(\omega t +i(k_x x + k_y y + k_z z))$ for the perturbation:

\begin{eqnarray}
k_x \delta u + k_y \delta v + k_z \delta w &=& 0 \cr
\omega \rho_0 \delta u + i k_x \delta P &=& 0\cr
\omega \rho_0 \delta v + i k_y \delta P &=& 0\cr
\omega \rho_0 \delta w + i k_z \delta P + \delta \rho g &=& 0\cr
\omega \delta X+ \delta w \frac{\partial X_0}{\partial z} - R_X \delta X-R_T\delta T &=&0 \cr
\omega \delta T + \delta w\left(\frac{\partial T_0}{\partial z}-\frac{\gamma-1}{\gamma}
\frac{T_0}{P_0}\frac{\partial P_0}{\partial z}\right)
 - H_X \delta X- H_T \delta T &=&0\cr
 \frac{\delta \rho}{\rho_0} + \frac{\delta T}{T_0} - \frac{\partial 
 \log \mu_0}{\partial X} \delta X &=& 0
\end{eqnarray}

We can eliminate directly $\delta u$ and $\delta v$ and define $1/h_p=-\partial \log P_0/\partial z$, $\nabla_T = -h_p \partial 
\log T_0/\partial z$, $\nabla_\mathrm{ad} = (\gamma-1)/\gamma$, 
and $k^2 = k_x^2+k_y^2+k_z^2$

\begin{eqnarray}
ik^2 \delta P +k_z \delta \rho g &=& 0 \cr
\omega \rho_0 \delta w + i k_z \delta P + \delta \rho g &=& 0\cr
\omega \delta X+\delta w\frac{\partial X_0}{\partial z} -  R_X\delta X- R_T\delta T&=&0\cr
\omega \delta T - \delta w \frac{T_0}{h_p} \left(\nabla_T-\nabla_\mathrm{ad}\right) -  H_X \delta X - H_T \delta T &=&0\cr
\frac{\delta \rho}{\rho_0}+\frac{\delta T}{T_0}-\frac{\partial\log \mu_0}{\partial X} \delta X &=&0
\end{eqnarray}

and we get a linear system for $\delta \rho$, $\delta X$, $\delta w$, $\delta T$, and 
$\delta P$, given by the matrix:

\begin{equation}
\left(\begin{array}{ccccc}
k_z g & 0 & 0 & 0 & ik^2 \\
0 & \omega-R_X & \frac{\partial X_0}{\partial z} & -R_T & 0\\
g & 0 & \omega \rho_0 & 0 & ik_z \\
0 & -H_X & -\frac{T_0}{h_p}\left(\nabla_T-\nabla_\mathrm{ad}\right) 
& \omega-H_T & 0 \\
-1/\rho_0 & \frac{\partial \log \mu_0}{\partial X} & 0 & -1/T_0 & 0
\end{array}\right)
\end{equation}

\subsection{Schwarzschild criterion}
To derive the Schwarzschild criterion, we can eliminate $\delta X$ from the system and 
assume $H=0$ in an adiabatic environment. The matrix reduces to

\begin{equation}
\left(\begin{array}{cccc}
k_z g  & 0 & 0 & ik^2 \\
g  & \omega \rho_0 & 0 & ik_z \\
0  & -\frac{T_0}{h_p}\left(\nabla_T-\nabla_\mathrm{ad}\right) & 
\omega & 0 \\
-1/\rho_0 & 0 & -1/T_0 & 0
\end{array}\right)
\end{equation}
whose determinant is given by 
\begin{eqnarray}
k_z g ik_z \frac{g}{h_p}(\nabla_T-\nabla_\mathrm{ad}) - 
 ik^2\frac{g}{h_p}(\nabla_T-\nabla_\mathrm{ad}) + ik^2\omega^2 = 0 \cr
\omega^2  -(\nabla_T-\nabla_\mathrm{ad})\frac{k_x^2+
k_y^2}{k^2}\frac{g}{h_p} = 0
\end{eqnarray}

This equation admits a positive real solution (which corresponds to the onset of adiabatic convective instability) if and only if $\nabla_T-\nabla_\mathrm{ad}>0$.

\subsection{Ledoux criterion}
We derive the Ledoux criterion for convection assuming an adiabatic environment $H=0$, 
a non-reactive/diffusive fluid $R=0$. 

The determinant of the matrix is given by 
\begin{eqnarray}
-i(k_x^2+k_y^2) \omega  \frac{g}{h_p}(\nabla_T-\nabla_\mathrm{ad}) 
+ ik^2 \omega^3 - i(k_x^2+k_y^2)\frac{\partial \log \mu_0}{\partial z} g \omega = 0 \cr
\omega^2  - \frac{g}{h_p}(\nabla_T-\nabla_\mathrm{ad}-\nabla_\mu)
\frac{k_x^2+k_y^2}{k^2} = 0 
\end{eqnarray}

with $\nabla_\mu = -h_p\partial \log \mu_0/\partial z$ and we get the Ledoux 
criterion for convective instabilities $\nabla_T-\nabla_\mathrm{ad} - \nabla_\mu>0$.
So a region unstable to the Schwarzschild criterion, can be stabilized by a $\mu$ gradient
 if $\nabla_\mu > \nabla_T-\nabla_\mathrm{ad}>0 $.

\subsection{Generalization for thermo-compositional convection}
With $R_{X,T} \neq 0$ and $ H_{T,X} \neq 0$, the determinant now is given by a third 
degree polynom $\omega^3 + a_2 \omega^2 + a_1\omega + a_0 = 0$ with

\begin{eqnarray}\label{eq:poly}
a_2 &=& -R_X-H_T \cr
a_1 &=& H_T R_X -H_X R_T-\frac{k_x^2+k_y^2}{k^2}\frac{g}{h_p}
(\nabla_T-\nabla_\mathrm{ad}-\nabla_\mu) \cr
a_0 &=&  \frac{k_x^2+k_y^2}{k^2}\frac{g}{h_p}((\nabla_T-
\nabla_\mathrm{ad})R_X-\nabla_\mu H_T) \cr
    &+& \frac{k_x^2+k_y^2}{k^2} T_0 \frac{g}{h_p}(\nabla_T-
    \nabla_\mathrm{ad}) \frac{\partial \log \mu_0}{\partial X}R_T\cr
    &+& \frac{k_x^2+k_y^2}{k^2} g \frac{1}{T_0} H_X\frac{\partial X_0}{\partial z}
\end{eqnarray}

According to the Hurwitz criterion \citep[e.g.][]{1966PASJ...18..374K}, one of the root of the polynom has a 
positive real part if at least one of the coefficients if negative (when they are all 
non-zero) $a_0 < 0$, $a_1 < 0$, or $a_2< 0$.
The conditions $a_2 < 0$, $a_1< 0$ lead to the following inequalities:

\begin{eqnarray}\label{eq:ledoux}
 -R_X-H_T &<& 0\cr
\nabla_T-\nabla_\mathrm{ad}-\nabla_\mu &>& (H_T R_X-H_X R_T) \frac{k^2}{k_x^2+k_y^2}\frac{h_p}{g}
\end{eqnarray}

For realistic physical conditions $R_X<0$ and $H_T<0$, hence the first inequality is never
met. The second inequality reduces to the Ledoux criterion because when $\nabla_T-\nabla_
\mathrm{ad}-\nabla_\mu>0$ we can always find wavenumbers $(k_x,k_y,k_z)$ 
that will satisfy the inequality. Note that if all the coefficients are positive, 
we can still have an instability if $a_2 a_1<a_0$ according to Hurwitz criterion, the interpretation of this instability condition remains an open question.

The instability condition $a_0<0$ leads to the following inequality:

\begin{equation}
(\nabla_T-\nabla_\mathrm{ad})\omega_X^\prime-\nabla_\mu\omega_T^\prime <0 
\end{equation}

with
\begin{eqnarray}
\omega_X^\prime &=& R_X + T_0 R_T \frac{\partial \log \mu_0}{\partial X}\cr
\omega_T^\prime &=&  H_T+  \frac{1}{T_0}H_X\left(\frac{\partial \log \mu_0}{\partial X} 
\right)^{-1}
\end{eqnarray}

This criterion describes the diabatic instability linked to the presence of source terms and is
the one that encompass thermohaline/fingering convection and moist convection. Note that the instability criterion remains well defined in the limit $\partial \log \mu_0/\partial X=0$. In that case, the inequality tends to:

\begin{equation}
(\nabla_T-\nabla_\mathrm{ad})R_X-\nabla_X X _0 H_X/T_0 <0 
\end{equation}

We can approximate also the growth rate of the instability by assuming that the polynom reduce to $a_1\omega+a_0 \sim 0$ for the diabatic instability in the limit of small growth rates when $a_1>0$ and $a_0<0$. We get then an approximate formula for the growth rate for $k_z<<k_{x,y}$:

\begin{equation}
\omega\sim \frac{\nabla_\mu \omega^\prime_T -
(\nabla_T-\nabla_\mathrm{ad})\omega^\prime_X}{(H_T R_X -H_X R_T)h_p/g-(\nabla_T-\nabla_\mathrm{ad}-\nabla_\mu)}
\end{equation}

When this approximation is not valid, the growth rate can be evaluated numerically directly from the dispersion relation.

\subsection{Moist convection criterion}

Let us assume that the reactive terms for composition are fast and at an equilibrium given 
by $X = X_\mathrm{eq} (P,T)$

The criterion can be written in the following form:
\begin{eqnarray}\label{eq:conv_eq}
\nabla_T-\nabla_\mathrm{ad} - \nabla_\mu \frac{\omega^\prime_T}{\omega^\prime_X}&>&0 \cr
\nabla_T-\nabla_\mathrm{ad} - \left(\frac{\partial\log P_0}{\partial z}\right)^{-1} 
\frac{\partial\log\mu_0}{\partial X}\frac{\partial X_\mathrm{eq}}{\partial z} 
\frac{\omega^\prime_T}{\omega^\prime_X}&>&0 \cr
\nabla_T-\nabla_\mathrm{ad} - \frac{\partial\log\mu_0}{\partial X}\left(\frac{\partial 
X_\mathrm{eq}}{\partial \log T}\nabla_T+\frac{\partial X_\mathrm{eq}}{\partial \log P} 
 \right)\frac{\omega^\prime_T}{\omega^\prime_X}&>&0 \cr
\nabla_T-\nabla_\mathrm{ad}\frac{1+ \frac{1}{\nabla_\mathrm{ad}}\frac{\partial\log\mu_0}
{\partial X}\frac{\partial X_\mathrm{eq}}{\partial \log P}\frac{\omega^\prime_T}
{\omega^\prime_X}}{1- \frac{\partial\log\mu_0}{\partial X}\frac{\partial X_\mathrm{eq}}
{\partial \log T}\frac{\omega^\prime_T}{\omega^\prime_X}} >0 &&
\end{eqnarray}
assuming that $\omega^\prime_X<0$ and $1- \frac{\partial\log\mu_0}{\partial X}
\frac{\partial X_\mathrm{eq}}{\partial \log T}\frac{\omega^\prime_T}{\omega^\prime_X}>0$ (always verified for moist convection).

We assume that the reaction source term $R$ is condensation/evaporation of water and the thermal source term is the corresponding release or pumping of latent heat $L$ and we neglect the temperature dependance of the latent heat. In that case the criterion becomes:
\begin{eqnarray}\label{eq:moist}
\omega^\prime_X &=& R_X +T_0 R_T \frac{\partial \log \mu_0}{\partial X}\cr
\omega^\prime_T &=& -\frac{1}{T_0}\left(\frac{\partial \log \mu_0}{\partial X}
\right)^{-1}\frac{\omega_X^\prime L}{c_p} \cr
\nabla_T &-&\nabla_\mathrm{ad}\frac{1- \rho_0\frac{\partial X_\mathrm{eq}}{\partial P}L}
{1 +\frac{\partial X_\mathrm{eq}}{\partial T}\frac{L}{c_p}} >0 
\end{eqnarray}
For Earth atmosphere, we can then express $X_\mathrm{eq}(P,T)$ as a function of vapour pressure 
at saturation $f(T)$ and $R_v$ and $R_d$ the vapour and dry air gas constant:
\begin{eqnarray}
X_\mathrm{eq}(P,T) &\approx& \frac{R_d}{R_v}\frac{f(T)}{P}\cr
f(T) &=& 6.11 \exp \left(\frac{L}{R_v}\left(\frac{1}{273}-\frac{1}{T}\right)\right)\quad \mathrm{hPa}\cr
\frac{\partial X_\mathrm{eq} }{\partial T}&\approx& \frac{L X_\mathrm{eq}}{R_v T^2}\cr
\frac{\partial X_\mathrm{eq} }{\partial P}&\approx& -\frac{X_\mathrm{eq}}{\rho R_d T}
\end{eqnarray}
which gives the standard criterion for Earth moist convection \citep[see][]{Stevens_2005}
\begin{equation}
\nabla_T -\nabla_\mathrm{ad}\frac{1+ \frac{X_\mathrm{eq}L}{R_d T_0}}
{1 +\frac{X_\mathrm{eq}L^2}{c_pR_v T_0^2}} >0
\end{equation}

\bibliographystyle{aasjournal}
\bibliography{main.bib}

\end{document}